\documentclass[12pt]{article}
\usepackage{amssymb,slashed}
\usepackage{amsmath}
\usepackage[all]{xy}
\newcommand{\ft}[2]{{\textstyle\frac{#1}{#2}}}
\def\Re{\mathop{\rm Re}\nolimits}
\def\Im{\mathop{\rm Im}\nolimits}
\def\Tr{\mathop{\rm Tr}\nolimits}
\def\tr{\mathop{\rm tr}\nolimits}
\def\rme{{\mathrm e}}
\def\rmi{{\mathrm i}}
\newcommand{\ii}{\mathrm{i}}
\def\rmd{{\mathrm d}}
\newsavebox{\uuunit}
\sbox{\uuunit}
    {\setlength{\unitlength}{0.825em}
     \begin{picture}(0.6,0.7)
        \thinlines
        \put(0,0){\line(1,0){0.5}}
        \put(0.15,0){\line(0,1){0.7}}
        \put(0.35,0){\line(0,1){0.8}}
       \multiput(0.3,0.8)(-0.04,-0.02){12}{\rule{0.5pt}{0.5pt}}
     \end {picture}}
\newcommand {\unity}{\mathord{\!\usebox{\uuunit}}}
\newcommand{\abs}[1]{\lvert#1\rvert}

\csname @addtoreset\endcsname{equation}{section}
\newcommand{\SU}{\mathop{\rm SU}}

\newcommand{\U}{\mathop{\rm {}U}}
\newcommand{\USp}{\mathop{\rm {}USp}}

\newcommand{\Sp}{\mathop{\rm {}Sp}}

\newcommand{\fsu}{\mathfrak{su}}
\newcommand{\fsp}{\mathfrak{sp}}
\newcommand{\fusp}{\mathfrak{usp}}
\newcommand{\rSp}{\mathrm{Sp}}
\newcommand{\rSU}{\mathrm{SU}}
\newcommand{\rU}{\mathrm{U}}

\newif\ifpdf
\ifx\pdfoutput\undefined
   \pdffalse
   \usepackage{cite}
 \else
   \pdfoutput=1
   \pdftrue
  \usepackage[pdftex]{hyperref}
\fi

\def\cD{{\cal D}}
\def\cK{{\cal K}}
\def\cL{{\cal L}}
\def\cN{{\cal N}}
\def\cP{{\cal P}}
\def\cQ{{\cal Q}}

\def\cV{{\cal V}}

\begin{document}

\begin{titlepage}
\begin{flushright}
KUL-TF-05/24\\
hep-th/0511001
\end{flushright}
\vspace{.5cm}
\begin{center}
\baselineskip=16pt
{\LARGE  $D$-term cosmic strings from $N=2$ Supergravity  
}\\
\vfill
{\large Ana Ach{\'u}carro $^{1,2}$, Alessio Celi$^3$, Mboyo Esole$^1$, \\
\vskip 0.2cm Joris Van den Bergh$^3$ and Antoine Van Proeyen$^3$
  } \\
\vfill
{\small $^1$ Lorentz Institute of Theoretical Physics, Leiden
University,\\ 2333 RA Leiden, The Netherlands \\ \vspace{6pt}
 $^2$
Department of Theoretical Physics, University of the Basque Country\\
UPV-EHU, 48080 Bilbao, Spain\\ \vspace{6pt}
$^3$ Instituut voor Theoretische Fysica, Katholieke Universiteit Leuven,\\
       Celestijnenlaan 200D B-3001 Leuven, Belgium.
      \\[2mm] }
\end{center}
\vfill
\begin{center}
{\bf Abstract}
\end{center}
{\small
 We describe new half-BPS cosmic string solutions in $N=2$, $d=4$
supergravity coupled to one vector multiplet and one hypermultiplet. They
are closely related to $D$-term strings in $N=1$ supergravity. Fields of
the $N=2$ theory that are frozen in the solution contribute to the
triplet moment map of the quaternionic isometries and leave their trace
in $N=1$ as a constant Fayet-Iliopoulos term. The choice of $\U(1)$
gauging and of special geometry are crucial. The construction gives rise
to a non-minimal K{\"a}hler potential and can be generalized to higher
dimensional quaternionic-K{\"a}hler manifolds.
 }\vspace{2mm} \vfill \hrule
width 3.cm
 {\footnotesize \noindent
e-mails: \{achucar,esole\}@lorentz.leidenuniv.nl,\\
\phantom{e-mails: } \{alessio.celi, joris.vandenbergh,
antoine.vanproeyen\}@fys.kuleuven.be }
\end{titlepage}
\addtocounter{page}{1}
 \tableofcontents{}
\newpage
\section{Introduction}

In this paper we discuss the classical embedding of $D$-term string
solutions of $d=4$, $N=1$ supergravity into $N=2$ theories. $D$-term
strings in supergravity \cite{Dvali:2003zh,Morris:1997ua} are
BPS-solutions of the supersymmetric Einstein-Higgs Abelian gauge field
model coupled to supergravity, with half of the supersymmetries unbroken,
saturating the Bogomol'nyi bound. Earlier work on BPS-strings and the
Bogomol'nyi bound  in $d=3$ supergravity can be found in
\cite{Becker:1995sp,Edelstein:1996ba,Edelstein:1995md}. Further work on
the properties of $D$-term strings includes the analysis of zero  modes
\cite{Holman:1997ey,Davis:1997bs,Jeannerot:2004bt} and BPS axionic
strings \cite{Binetruy:1998mn,Davis:2005jf,Blanco-Pillado:2005xx}.
$D$-term strings are expected to form after $D$-term inflation
\cite{combDterminfl};  a recent assessment of their impact on the Cosmic
Microwave Background can be found in \cite{Rocher:2004my} and references
therein. The supergravity model considered in \cite{Dvali:2003zh}
contains one vector-multiplet, one chiral multiplet and the graviton
multiplet; the model has a complex scalar, charged under $\U(1)$, that
parametrizes the trivial (flat) internal K{\"a}hler-Hodge space with K{\"a}hler
potential ${\cal K}=\phi \phi^*$, a $D$-term potential, and a vanishing
superpotential. Essential for the construction is the constant
\emph{Fayet-Iliopoulos term} (henceforth denoted FI term
\cite{Fayet:1974jb}) appearing in the $D$-term potential; issues
concerning this term are clarified in
\cite{Binetruy:2004hh,VanProeyen:2004xt}.

Engineering such a constant FI term from $N=2$ supergravity is not
trivial: an FI term in $N=2$ supergravity corresponds to the case of an
arbitrary constant in the \emph{moment map}, which can only occur when
there are no hypermultiplets. However, one needs at least one
hypermultiplet to play the role of matter charged under the $\U(1)$
gauging that we intend to perform (scalars of vector multiplets cannot be
charged under an Abelian gauging). To perform the gauging, one chooses a
suitable isometry of the (quaternionic-K{\"a}hler) hypermultiplet target
space. The moment map corresponding to the isometry is then dependent on
some or all of the hypermultiplet scalars (see \cite{Andrianopoli:1997cm}
for a review). In this paper we propose an ansatz that circumvents this
problem. Formally, the ansatz is closely related to a \emph{consistent
truncation} of $N=2$ to $N=1$, such that the cosmic string solutions
would be half-BPS $D$-term strings in the $N=1$ theory, with (constant)
FI term and vanishing superpotential. The interesting point here is that
we show they are also half-BPS in the $N=2$ theory.

The authors of \cite{Dvali:2003zh} recently conjectured that $D$-term
strings might be the low energy manifestation of D1 or wrapped D$p$
branes (see \cite{Dvali:2003zj,Copeland:2003bj} and the recent review
\cite{Majumdar:2005qc}), based on the observation that $D$-term strings
are the only half-BPS vortices available in $N=1$ supergravity. Support
for this conjecture was provided in
\cite{Halyo:2003uu,Lawrence:2004sm,Matsuda:2004qt}. Since then, it has
been shown that other $D$-term BPS vortices (axisymmetric solutions)
exist in $N=1$ supergravity (for instance semilocal
``vortices'' in the Bogomol'nyi limit
\cite{Urrestilla:2004,Dasgupta:2004dw}, which have arbitrarily wide
cores, see also \cite{Chen:2005ae}), an observation that may be relevant
to the conjecture. In any case, the idea of matching the BPS D-brane
states of superstring theory with some low-energy counterparts in
supergravity is a very interesting one. If such a correspondence can be
made, then one would in principle expect to find half-BPS vortices also
in other low-energy manifestations, in particular in compactifications
with $N=2$. Our results show that this expectation is correct.

Consistent truncations of $N=2$ to $N=1$ are described in
\cite{Andrianopoli:2001zh,Andrianopoli:2001gm}, where the authors
consider the truncated $N=2$ BPS equations in the reduced $N=1$ theory
and analyse the resulting geometric conditions on the special K{\"a}hler and
quaternionic-K{\"a}hler target spaces. We rely heavily on these results to
show the consistency of the ansatz. The other ingredients are the choice
of $\U(1)$-isometry to be gauged and the choice of special geometry,
which are essential to the construction. We give two explicit examples of
``minimal'' $N=2$ models with one vector- and one hypermultiplet. They have
special K{\"a}hler space $\SU(1,1)/\U(1)$ and a symmetric one-dimensional
quaternionic-K{\"a}hler space, for which there are two choices, both reducing
after truncation to the same $N=1$ action (up to a normalization) with
one chiral and one vector multiplet. We also comment on the embedding in
models with more than one hypermultiplet.

The $N=1$ action resulting from the truncation has a complex scalar that
parametrizes the K{\"a}hler-Hodge manifold $\SU(1,1)/\U(1)$. We solve the BPS
equations and we describe the cosmic string half-BPS solution, in close
analogy to the results in \cite{Dvali:2003zh}. As an aside, we use a
Bogomol'nyi-type argument to prove that cylindrically symmetric $N=1$
$D$-term strings with complex scalars parametrizing any K{\"a}hler-Hodge
manifold, and charged under an Abelian $\U(1)$, have a mass per unit
length that is bounded below by the Gibbons Hawking surface term, and
that the bound is attained by the BPS solutions (if these exist). Since
we have assumed a specific ansatz, we cannot immediately conclude that
the solutions are stable but at least they minimize the energy within
this class of configurations.

We begin in section \ref{ss:basicN2} by reviewing some basic ingredients
of $N=2$ supergravity that we will use, see also
\cite{Andrianopoli:1997cm}. It includes the kinetic terms of the
Lagrangian and the potential, as well as the leading terms of the
supersymmetry transformations. We remind the reader of the problem
concerning constant FI terms in $N=2$ supergravity in section
\ref{ss:isomMomMap}. The mechanism by which FI terms in $N=1$ originate
from $N=2$ theories, and its relation to consistent truncations, is
explained in section \ref{ss:FIconsistentTrunc}. It is illustrated there
on the 1-dimensional quaternionic-K{\"a}hler symmetric spaces. In section
\ref{ss:TruncN2BPS}, we first define the special geometry that is used in
the models that we consider. A field configuration is then presented with
the property that the bosonic action becomes just the one of an $N=1$
matter-coupled supergravity system, and that the BPS equations decouple.
This resulting system is studied further in section \ref{ss:resultN1},
where we introduce an ansatz for a cosmic string. Two different
parametrizations of the 1-dimensional complex projective space are
convenient in different settings. We compare them, and after splitting
the conditions, the BPS equations can be solved. From the $N=2$ point of
view, the relation between the half-plane and the unit disk is an
$\SU(2)$ rotation of the quaternionic structure. In the $N=1$ reduced
theory, such an $\SU(2)$-connection is seen as a K{\"a}hler transformation or
better as a transformation of the K{\"a}hler $\U(1)$-connection. As this is
an example of $D$-term strings with non trivial K{\"a}hler-Hodge target
spaces (see also \cite{Blanco-Pillado:2005xx}),
we give in section \ref{s:N=1Sugra} some general remarks on such $N=1$ string solutions.
Section \ref{ss:conclusions} gives our conclusions.

We have included an appendix \ref{app:notation} with notations, and an
appendix \ref{app:paramCosets} to expose the parametrizations that we use
for the coset spaces.

\section{Basic formulae of matter-coupled $N=2$ supergravity}
\label{ss:basicN2}

\subsection{Fields and kinetic terms}
We repeat here the basic formulae of $N=2$ supergravity coupled to $n_V$
vector multiplets and $n_H$ hypermultiplets
\cite{deWit:1984pk,deWit:1985px,D'Auria:1991fj,Andrianopoli:1997cm},
though in the following sections we will mostly use $n_V=n_H=1$. These
theories contain the fields given in table \ref{tbl:fields}.
\begin{table}[htbp]
\caption{\it Multiplets and fields of the super-Poincar{\'e}
theories}\label{tbl:fields}
\begin{center}
 \begin{tabular}{|l|c|l|}
\hline
 vielbein &  $e_\mu ^a$  & $\mu ,a=0,\ldots ,3$ \\
 gravitini &   $\psi _\mu ^i$, $\psi _{\mu i}$ &  $i=1,2$  \\
 vectors & $W_\mu ^I$ & $I=0,\ldots ,n_V$ \\
 gaugini  &  $\lambda ^\alpha _i$, $\lambda ^{\bar \alpha i}$ & $\alpha =1,\ldots ,n_V$ \\
 hyperini & $\zeta ^A $ , $\zeta _A$& $A=1,\ldots ,2n_H$ \\
 K{\"a}hler manifold scalars & $z^\alpha $, $\bar z^{\bar \alpha }$ & \\
 hyperscalars &  $q^X$ & $X=1,\ldots ,4n_H$ \\
\hline
\end{tabular}\end{center}
\end{table}
For the fermions, we write the left-handed components [projected by
$\ft12(1+\gamma _5)$] on the left-hand side and the right-handed
component on the right-hand side. The vector multiplets contain the
complex\footnote{For the quantities in the K{\"a}hler manifolds, we use the
bar notation for complex conjugation. In the hypermultiplets we
distinguish hermitian conjugation indicated by the bar, and complex
conjugation indicated by $*$. Charge conjugation, which is complex
conjugation for bosons, replaces left-handed fermions with right-handed
ones.} scalars $z^\alpha $, while the scalars for the hypermultiplets are
written here as real fields $q^X$. The leading (kinetic) terms of the
action are then
\begin{eqnarray}
  e^{-1}{\cal L}_{\rm kin}&=&\ft12 R-\bar \psi _\mu ^i\gamma ^{\mu \nu \rho }\nabla _\nu \psi_{\rho i}
  +\ft12\Im \left( {\cal N}_{IJ}F_{\mu \nu }^{+I}F^{+\mu \nu J}\right)
   -\ft12g_{\alpha \bar \beta }\bar \lambda
  ^\alpha _i\slashed{\nabla }\lambda ^{\bar \beta i}
  -g_{\alpha \bar \beta }\nabla_\mu z^\alpha \nabla ^\mu \bar z^{\bar \beta }\nonumber\\
  &&-\ft12 g_{XY}\nabla _\mu q^X\nabla ^\mu q^Y
  -2\bar \zeta ^A\slashed{\cal D}\zeta _A .
 \label{L4leading}
\end{eqnarray}
See Appendix \ref{app:notation} for metric and spinor conventions and  $e
= \det (e_\mu^ a)$. The derivatives $\nabla $ are in this approximation
ordinary spacetime derivatives, but are in the full theory covariant
derivatives that we will explain below, see (\ref{covder}). Here $F_{\mu
\nu }^{+I}$ is the self-dual combination
\begin{equation}
  F_{\mu \nu }^{\pm I}=\ft12\left( F^I_{\mu \nu }\mp \ft12 \rmi e\varepsilon_{\mu \nu \rho \sigma
  }F^{I\rho\sigma } \right) ,\qquad F^I_{\mu \nu }=2\partial _{[\mu }W_{\nu
  ]}^I=\partial _\mu W_\nu ^I-\partial _\nu W_\mu ^I,
 \label{defF+-}
\end{equation}
where we restrict ourselves to the Abelian case. The quantities ${\cal
N}_{IJ}$, $g_{\alpha \bar \beta }$ and $g_{XY}$ are related to the chosen
geometry for the vector multiplets and the hypermultiplets, which we will
now describe.

The vector multiplet action describes a special K{\"a}hler manifold.
Everything is determined in terms of a prepotential\footnote{A more
general description without a prepotential is possible
\cite{Ceresole:1995jg,Craps:1997gp}, but is not needed here.}, $F(Z)$,
which should be holomorphic and homogeneous of second degree in variables
$Z^I$. The basic object is a $2(n_V+1)$-component symplectic section
\begin{eqnarray}
&&  V(z,\bar z)=\begin{pmatrix}X^I\cr M_I\end{pmatrix}=
  \rme^{{\cal K}(z,\bar z)/2}v(z),\qquad v(z)=\begin{pmatrix}Z^I(z)\cr F_I
\end{pmatrix},\nonumber\\ && M_I=\frac{\partial }{\partial X^I}F(X),\qquad
 F_I=\frac{\partial }{\partial Z^I}F(Z).
\label{Vlocal}
\end{eqnarray}
Here $Z^I(z)$ are arbitrary functions (up to conditions for
non-dege\-nera\-cy), reflecting the freedom of choice of coordinates
$z^\alpha $. The lower components depend on the prepotential. A
constraint in terms of a symplectic inner product
\begin{equation}
  <V,{\bar V}>=X^I\bar M_I-M_I\bar X^I=\rmi,
\label{VbarVisi}
\end{equation}
determines the K{\"a}hler potential as
\begin{equation}
   \rme^{-{\cal K}(z,\bar z)}=-\rmi \langle {v},{\bar
  v}\rangle =-\rmi Z^I \bar F_I +\rmi F_I\bar Z^I.
 \label{Kahlerpot}
\end{equation}
The metrics for the scalars and for the vectors are then determined by
\begin{equation}
  g_{\alpha \bar \beta }=\partial _\alpha \partial _{\bar \beta } {\cal
  K}(z,\bar z)=\rmi \langle {\cal D}_\alpha v,{\cal D}_{\bar \beta }\bar
  v\rangle,\qquad
{\cal N}_{IJ}\equiv \left( \begin{array}{cc}F_I &\bar {\cal
D}_{\bar\alpha}\bar F_I
\end{array}\right)  \left( \begin{array}{cc}Z^J &\bar
{\cal D}_{\bar\alpha}\bar Z^J
\end{array}\right)^{-1}  ,  \label{sympldefgcN}
\end{equation}
where covariant derivatives are defined by
\begin{equation}
  {\cal D}_\alpha v=\partial _\alpha v+(\partial _\alpha {\cal K})v,\qquad
{\cal D}_{\bar \alpha} \bar v=\partial_{\bar \alpha}\bar
v+(\partial_{\bar \alpha} {\cal K})\bar v.
 \label{defKacovder}
\end{equation}
Due to the presence of the prepotential, one can give also the
expressions
\begin{eqnarray}
&&  \rme^{-{\cal K}}=-Z^IN_{IJ}\bar Z^J,\qquad {\cal N}_{IJ}=\bar F_{IJ}
+\rmi \frac{N_{IN}N_{JK}\,Z^N Z^K}{N_{LM}\ Z^L Z^M},\nonumber\\ &&
N_{IJ}\equiv 2\Im F_{IJ}= -\rmi F_{IJ}+\rmi \bar F_{IJ},\qquad
F_{IJ}=\frac{\partial^2}{\partial Z^I\partial Z^J} F(Z).
 \label{NdefResult}
\end{eqnarray}
A useful relation between the two metrics is
\begin{equation}
  -\ft12(\Im{\cal N})^{-1|IJ}=\mathcal{D}_\alpha X^I g^{\alpha \bar \beta}\mathcal{D}_{\bar \beta }\bar
  X^J + \bar X^IX^J.
 \label{InverseImN}
\end{equation}

The hypermultiplets describe a quaternionic-K{\"a}hler manifold. The starting
point for a supergravity description is the vielbein $f_X^{iA}$. We need
furthermore a symplectic metric $\mathbb{C}_{AB}$, antisymmetric and with
complex conjugate $\mathbb{C}^{AB}$, such that it satisfies the same
relation as $\varepsilon _{ij}$:
\begin{equation}
  \mathbb{C}_{AC}\mathbb{C}^{BC}=\delta _A^B.
 \label{propC}
\end{equation}
The vielbein satisfies the reality property
\begin{equation}
  \left( f_X^{iA}\right) ^*=f_{XiA}=f_X^{jB}\varepsilon
  _{ji}\mathbb{C}_{BA}.
 \label{realvielbein}
\end{equation}
The inverse of the vielbein as $4n_H\times 4n_H$ matrix is $f^X_{iA}$.
The vielbein determines the metric and the quaternionic structures:
\begin{equation}
  2f_{X}^{jA}f_{YiA}=\delta _i^jg_{XY} +J_{XYi}{}^j,\qquad
  f^X_{iA}=g^{XY}f_{YiA}.
 \label{metricJfromf}
\end{equation}
$J_{XYi}{}^j$ is traceless in the $i,j$ indices, and is decomposed in the
3 complex structures as
\begin{equation}
J_{XYi}{}^j=\rmi (\sigma ^x)_i{}^j  J^x_{XY},\qquad  x=1,2,3,
 \label{complexstructuresx}
\end{equation}
where $\sigma ^x$ are the Pauli matrices. These complex structures are
covariantly constant using the Levi-Civita connection $\nabla ^{\rm LC}$
and an $\SU(2)$ connection $\omega _X^x$ as
\begin{equation}
  \nabla _X J^x_{YZ}\equiv \nabla ^{\rm LC}_XJ^x_{YZ}+2\varepsilon ^{xyz}\omega
  _X^y J^z_{YZ}=0.
 \label{nablaJ}
\end{equation}
The complex structure is proportional to the $\SU(2)$ curvature. Written
as forms, this relation is
\begin{equation}
  {\cal R}^x\equiv \rmd \omega ^x +\varepsilon ^{xyz}\omega ^y\omega ^z= \ft12\nu
  J^x, \qquad \nu =-\kappa ^2=-1.
 \label{RSU2isJ}
\end{equation}
The value of $\nu $ is arbitrary in quaternionic-K{\"a}hler manifolds, but
invariance of the action in supergravity relates it to the gravitational
coupling constant, which we have put equal to~1 in this paper.

\subsection{Isometries and the moment map}
\label{ss:isomMomMap}

As we will consider only an Abelian vector multiplet, we will restrict
this presentation to the gauging of isometries in the hypermultiplet. We
consider the transformation with parameters $\alpha ^\Lambda$:
\begin{equation}
  \delta _G q^X= -g\alpha ^\Lambda k_\Lambda^X,
 \label{transfoGq}
\end{equation}
and $k_\Lambda^X$ are Killing vectors. In a quaternionic-K{\"a}hler manifold
any isometry normalizes\footnote{This means that the Lie derivative of
the three complex structures is a linear combination of the complex
structures themselves.} the quaternionic structure
\cite{Alekseevsky:1993}. The Killing vector can then be derived from a
triplet moment map $\cP_\Lambda^x$:
\begin{equation}
\iota_\Lambda J^x \equiv k_\Lambda ^X J^x_{XY}\rmd q^Y=2\nabla
\cP_\Lambda^x\equiv
2(\rmd\cP^x_\Lambda+2\varepsilon^{xyz}\omega^{y}\cP^z_\Lambda) \qquad
\mbox{or}\qquad 4n_H{\cal P}^x_\Lambda =J^{xXY}\partial _Xk_{Y\Lambda }.
\label{iJ}
\end{equation}
Due to the non-trivial $\rSU(2)$ connection, the triplet moment maps
cannot be shifted by arbitrary constants, (unlike in rigid $N=2$
supersymmetry, where only $\rmd {\cal P}^x$ occurs in $\iota_\Lambda
J^x$). These constants would be the FI terms, and for the above reasons
their introduction in $N=2$ supergravity is problematic.

The moment map can also be described in another way. A Killing vector
preserves the connection $\omega^x$ and K{\"a}hler two forms $J^x$ only
modulo an $\rSU(2)$ rotation. Denoting by $\cL_\Lambda $ a Lie derivative
with respect to $k_\Lambda $, we have
\begin{equation}
\cL_\Lambda \omega^x  = -\ft12 \nabla r^x_\Lambda ,\quad \cL_\Lambda
J^x=\varepsilon^{xyz}r^y_\Lambda J^z, \label{Liederinr}
\end{equation}
where $r^x_\Lambda$ is known as an $\rSU(2)$ {\em compensator}. The
$\rSU(2)$-bundle of a quaternionic manifold is non-trivial and therefore
it is  impossible to get rid of the compensator $r_\Lambda^x$ by a
redefinition of the $\rSU(2)$ connections.\footnote{Again, this is in
contrast with $N=2$ rigid supersymmetry, since hyper-K{\"a}hler manifolds
have a trivial $\rSU(2)$ bundle, and therefore no compensator.} The
moment map can be expressed in terms of the triplet of connections
$\omega^x$ and the compensator $r^x_\Lambda$ in the following way
\cite{Galicki:1987ja}:
\begin{equation}
\cP^x_\Lambda =\ft12 r^x_\Lambda+\iota_\Lambda \omega^x.
\label{momentmap}
\end{equation}

\subsection{Gauging and supersymmetry transformations}

We now gauge some of the isometries mentioned above, and connect them to
gauge transformations of the vectors, such that the indices $\Lambda $
are replaced by $I$ with
\begin{equation}
  \delta _G W_\mu ^I= \partial _\mu \alpha ^I.
 \label{gaugeVector}
\end{equation}
This modifies the supersymmetry transformation laws. The normalizations
are such that the bosons transform as
\begin{eqnarray}
 \delta e_\mu ^a& = & \ft12\bar \epsilon ^i\gamma ^a\psi _{\mu i}+\ft12\bar \epsilon _i\gamma ^a\psi _\mu ^i, \nonumber\\
\delta W_\mu ^I&=&\ft12({\cal D}_\alpha X^I)\varepsilon^{ij} \bar
\epsilon _i\gamma_\mu\lambda _j^\alpha +\ft12({\cal D}_{\bar \alpha} \bar
X^I)\varepsilon_{ij} \bar \epsilon ^i\gamma_\mu\lambda ^{\bar \alpha j}+
\varepsilon ^{ij}\bar \epsilon _i\psi _{\mu j} X^I+ \varepsilon
_{ij}\bar \epsilon ^i\psi^j _\mu \bar  X^I, \nonumber\\
 \delta z^\alpha  & = & \ft12\bar \epsilon ^i\lambda _i^\alpha , \nonumber\\
\delta q^X & = & -\rmi f^X_{iA}\bar \epsilon ^i\zeta ^A+\rmi f^{XiA}\bar
\epsilon _i\zeta _A.
 \label{susybosons}
\end{eqnarray}
For a bosonic configuration, the $N=2$ supersymmetry transformations of
the left-handed fermionic fields are (see appendix \ref{app:notation} for
a description of our conventions):
\begin{eqnarray}
 \delta \psi _\mu ^i & =  & \nabla _\mu(\omega ) \epsilon^i
 -g\gamma_\mu S^{ij}\epsilon _j
 +\ft14 \gamma^{\rho \sigma }T^-_{\rho \sigma }\varepsilon^{ij}\gamma_\mu\epsilon_j,\nonumber\\
 \delta \lambda ^\alpha_ i&=&\slashed{\nabla }z^\alpha \epsilon _i -\ft12g^{\alpha \bar \beta }{\cal D}_{\bar \beta }
 \bar X^I\Im{\cal N}_{IJ}F^{-J}_{\mu \nu }\gamma ^{\mu \nu }\varepsilon _{ij}\epsilon^j
+ gN^\alpha _{ij}\epsilon ^j,\nonumber\\
 \delta \zeta ^A&=&\ft12\rmi f^{Ai}_X\slashed{\nabla }q^X\epsilon_i
+g{\cal N}^{iA}\varepsilon _{ij}\epsilon ^j.
\label{N=2susyeqns}
\end{eqnarray}
The covariant derivatives are
\begin{eqnarray}
 \nabla _\mu(\omega) \epsilon^i &\equiv &
 \left( \partial_\mu + \ft14 \omega_\mu{}^{ab} \gamma _{ab}\right) \epsilon ^i
 +\ft12\rmi A_\mu \epsilon^i+ V_{\mu j}{}^i\epsilon ^j,\nonumber\\
 \nabla _\mu z^\alpha &= & \partial _\mu z^\alpha +gW_\mu ^I k_I^\alpha,  \nonumber\\
 \nabla _\mu q^X & = & \partial _\mu q^X+gW_\mu ^Ik_I^X.
 \label{covder}
\end{eqnarray}
We included here the effect of a gauging in the vector multiplet sector
by the Killing vector $k_I^\alpha $ describing the transformations under
the gauge symmetry of the vector multiplet scalar similar to the
definition of $k_I^X$ as in (\ref{transfoGq}) for the hypermultiplet
scalars. The $\SU(2)$ connection $V_{\mu i}{}^j$ is related to the
quaternionic-K{\"a}hler $SU(2)$:
\begin{equation}
 V_{\mu i}{}^j =\partial_\mu q^X \omega_{Xi}{}^j + gW_\mu^I
 {\cal P}_{Ii}{}^j.
 \label{SU2connection}
\end{equation}
$A_\mu $ are the components of the one-form gauge field of the K{\"a}hler
$\U(1)$:
\begin{equation}
  A=-\ft12\rmi \left( \partial _\alpha {\cal K}\rmd z^\alpha
  -\partial _{\bar \alpha }{\cal K}\rmd \bar z^{\bar \alpha }\right) .
 \label{valueAmu}
\end{equation}
In the case of gauging in the vector multiplet sector, this is modified
by a scalar moment map similar to the $\SU(2)$ connection. The dressed
graviphoton is given by
\begin{equation}
T^-_{\mu\nu}=F^{-I}_{\mu\nu}\Im{\cal N}_{IJ}X^J.
\end{equation}

The fermionic shifts (mass matrices) are given in terms of the
prepotentials and Killing vectors of the quaternionic-K{\"a}hler geometry
(dressed with special geometry data) as follows:
\begin{eqnarray}
S^{ij}&\equiv &-{\cal P}_I^{ij} X^I,\nonumber\\
N^\alpha _{ij}&\equiv &  \varepsilon _{ij}k^ \alpha _I \bar X^I -2{\cal
P}_{I ij}\bar {\cal D}_{\bar \beta }\bar X^I
  g^{\alpha \bar \beta } ,
 \qquad {\cal N}^{iA}\equiv -\rmi f^{iA}_Xk_I^X\bar X^I.
 \label{shifts}
\end{eqnarray}
They determine also the potential by
\begin{eqnarray}
  g^{-2}{\cal V}&=&-6S^{ij}S_{ij}+\ft12g_{\alpha \bar \beta }N^{\alpha }_{ij}N^{\bar \beta
  ij}+2{\cal N}^{iA}{\cal N}_{iA} \nonumber\\
&=&4\left ( U^{IJ}-3 \bar X^I X^J \right ) {\cal P}^x_I {\cal
P}^x_J+g_{\alpha \bar \beta}k^\alpha_I k^{\bar \beta}_J +2 g_{XY} k^X_I
k^Y_J \bar X^I X^J, \label{Vd4total}
\end{eqnarray}
where
\begin{equation}
  {U}^{IJ} \equiv g^{\alpha \bar \beta} {\cal D}_\alpha X^I
  {\cal D}_{\bar \beta} \overline{X}^J
=  -\ft{1}{2} \left( \mbox{Im}\mathcal{N}\right) ^{-1 \vert IJ}
-\overline{X}^I \, X^J. \label{formulaU}
\end{equation}

For the minimal models described in this work we will consider the case
of one vector multiplet, of which the scalar $z$ parametrizes
$\SU(1,1)/\U(1)$, and one hypermultiplet. A $\U(1)$ isometry of the
quaternionic-K{\"a}hler space with associated Killing vector $k_1$ will be
gauged with the vector $W^1$.

\section{FI terms from truncations of $N=2$ to $N=1$}
\label{ss:FIconsistentTrunc}

\subsection{General method}

If a reduction of $N=2$ to $N=1$ is performed by consistently truncating
the second gravitino $\psi_\mu^2$, the $N=1$ superpotential is a function
of $\cP^1+\ii\cP^2$ and the $D$-term is constructed out of $\cP^3$
\cite{Andrianopoli:2001zh}.

The aim now is to find a gauging that is consistent with the truncation
to $N=1$. The gauging should give rise to a moment map with
$\cP_1=\cP_2=0$ after truncation, and a non-zero component $\cP_3$ that
will result in a $D$-term potential. Thus, $\cP_3$ should contain a term
that acts as a FI term in the resulting $N=1$ theory. \textit{We can then
reinterpret the truncation as an ansatz, keeping both supersymmetries,
that allows us to solve the BPS equations of the full $N=2$ theory.} In
this way we find $D$-term solutions of $N=1$ with FI terms, embedded in
$N=2$ supergravity.

Equation (\ref{momentmap}) can be used to obtain a moment map with these
properties, which we illustrate by means of examples involving the two
normal quaternionic-K{\"a}hler manifolds of (quaternionic) dimension one.
This can be generalized to normal quaternionic manifolds of higher
dimension as they always contain as a completely geodesic submanifold one
of the two quaternionic manifolds of dimension one equipped with an
induced quaternionic structure
\cite{Alekseevsky1975,deWit:1992nm,Cortes1996}.

\subsection{Example of $\frac{\rSp(1,1)}{\rSp(1)\rSp(1)}$}

Details on the geometry and coset parametrization of the coset space
$\frac{\rSp(1,1)}{\rSp(1)\rSp(1)}$ are given in
appendix~\ref{app:paramCosets}. The space is characterized by the
following metric:
\begin{equation}
\rmd s^2=
 (\rmd h)^2+
\rme^{-2h}\left[{ (\rmd b^1)^2+(\rmd b^2)^2+(\rmd b^3)^2}\right].
 \end{equation}
Consider now the Killing vector\footnote{The Killing vector
(\ref{Killing1Sp1}) is one that rotates the quaternionic structure, while
its invariant subspace will define the truncated manifold. Furthermore,
it preserves the $J^3$ complex structure, which is the one that will play
the role of complex structure in the truncated, $N=1$, theory.}
\begin{equation}
k_1= 2b^1\frac{\partial}{\partial b^2}-2b^2\frac{\partial}{\partial b^1}.
\label{Killing1Sp1}
\end{equation}
It rotates the $\rSU(2)$ connection as follows:
\begin{equation}
\cL_{k_1} \omega^x =-2\varepsilon^{xyz} \omega^y \delta^z_3 = -\nabla
\delta^x_3,
\end{equation}
which implies that the compensator $r_{k_1}$ is a constant:  $r_{k_1}^x
=2\delta^x_3$. The  moment map can be computed as
\begin{equation}
\cP_{k_1} =\iota_{k_1} \omega  +\ft12r_{k_1} = \rme^{-h}
\begin{pmatrix}
 b^2 \\
-b^1  \\
0
\end{pmatrix}
+
\begin{pmatrix}
0 \\
0 \\
1
\end{pmatrix} .
\end{equation}
To have a vanishing superpotential, we impose the condition
$\cP^1=\cP^2=0$:
\begin{equation}
b^1= b^2 =0. \label{b1b20}
\end{equation}
This configuration defines a consistent truncation to a K{\"a}hler-Hodge submanifold of $\frac{\rSp(1,1)}{\rSp(1)\rSp(1)}$:
\begin{equation}
\xymatrix{ \frac{\rSp(1,1)}{\rSp(1)\rSp(1)} \ar[rr]^{b^1=b^2=0}
 &&
{\frac{\rSU(1,1)}{\U(1)} }. }
\end{equation}
One can check indeed that three of the ten isometries of the quaternionic
space are preserved by the truncation (the shift of $b^3$, the dilatation
coming from the Cartan generator, and one of the compact generators; the
latter is the generator that we will gauge), and that they form the
algebra $\SU(1,1)$. $\frac{\rSU(1,1)}{\U(1)}$ can be parametrized by the
complex field
\begin{equation}
\Phi=-b^3+\ii \rme^{h},
 \label{defPhiSp11}
\end{equation}
in terms of $\Phi$ the metric can be written as
\begin{equation}
\rmd s^2= \frac{\rmd\Phi\rmd\overline{\Phi}}{(\Im \Phi)^2}.
 \label{ds2Sp11}
\end{equation}
If we now use as Killing vector $\xi k_1$, with $\xi$  an arbitrary real
constant, the $N=1$ potential will have a constant Fayet-Iliopoulos term
given by the $D$-term
\begin{equation}
D=2g\cP^3=2 g\xi.
 \label{DinP3}
\end{equation}
We denote this quantity by $D$ as it is the $D$-term of $N=1$ (the
normalization will be explained in section \ref{sss:N=1BPS}). In this
way, an \emph{arbitrary} FI constant can be introduced in the $D$-term
potential of the reduced $N=1$ theory. In this reduced theory, $k_1$
identically vanishes so that it does not act on
$\frac{\rSU(1,1)}{\rU(1)}$. Therefore, the only effect of the gauging is
the generation of an FI term.

To generate a $D$-term potential with the right properties for $D$-term
string solutions, we  gauge a linear combination of $k_1$ and the
following one, which is the uplift of the compact isometry of
$\frac{\rSU(1,1)}{\U(1)}$:
\begin{equation}
k_2= 4 b^3\frac{\partial }{\partial h}+4b^1b^3\frac{\partial }{\partial
b^1}+4b^2b^3\frac{\partial }{\partial b^2} +
2\left[(b^3)^2-\rme^{2h}+1-(b^1)^2-(b^2)^2\right]\frac{\partial
}{\partial b^3},
\end{equation}
with corresponding moment map
\begin{equation}
\cP_{k_2}=\begin{pmatrix}
-2b^2-2b^1 b^3 \rme^{-h}\\
2b^1-2b^2 b^3 \rme^{-h}\\
-\rme^{-h}\left[ (b^3)^2  +1  -(b^2)^2 -(b^1)^2\right] -\rme^{h}
\end{pmatrix}.
\end{equation}
These Killing vectors automatically satisfy the requirements of
\cite{Andrianopoli:2001zh,Andrianopoli:2001gm}.

Gauging the linear combination $k=k_2+\xi  k_1$ and imposing the
truncation (\ref{b1b20}) then results in  an $N=1$ theory with vanishing
superpotential and $D$-term
\begin{equation}
D=-2g\left[  \rme^{-h} (b^3)^2 +\rme^{-h} +\rme^{h}\right] +2g\xi.
\end{equation}

\subsection{Example of $\frac{\rSU(2,1)}{\rSU(2)\U(1)}$}

The metric can be found using the solvable algebra approach (see appendix \ref{app:paramCosets}) or in the literature,
for example in the work of \cite{Ceresole:2001wi}. It is given by:
\begin{equation}
\rmd s^2= \ft{1}{2}\rmd h^2+\ft12\rme^{-2h}\left(\rmd b^3-e^1 \rmd
e^2+e^2 \rmd e^1\right)^2+ \rme^{-h}\left[(\rmd e^1)^2+(\rmd
e^2)^2\right].
\end{equation}
Killing vectors and moment maps can be found in \cite{Ceresole:2001wi}.
The correspondence with our notation is given in
(\ref{conventionCeresole}). Setting
\begin{equation}
e^1=e^2=0,
 \label{e1e20}
\end{equation}
the space can be consistently ``truncated'' to the submanifold
$\frac{\rSU(1,1)}{\U(1)}$ in the upper half-plane parametrization. We
define the complex field $\Phi$ by
\begin{equation}
\Phi=b^3+\ii \rme^{h},
 \label{defPhiSU21}
\end{equation}
leading to the metric differing from (\ref{ds2Sp11}) by a normalization
factor:
\begin{equation}
\rmd s^2= \frac{\rmd\Phi\rmd\overline{\Phi}}{2(\Im \Phi)^2}.
 \label{ds2SU21}
\end{equation}
This different normalization is due to another embedding of the
R-symmetry $\SU(2)$ in the two quaternionic-K{\"a}hler manifolds.

Along the lines of the previous section, we gauge a combination
\begin{equation}
  k=2\xi k_4 +2k_6 +2k_1,
 \label{kk4k6k1}
\end{equation}
following the labelling in \cite{Ceresole:2001wi}. The Killing vector in
the basis $(h,\,b^3,\,e^1,\,e^2)$ is
\begin{equation}
  k=2\xi \begin{pmatrix}0\cr 0\cr e^2\cr -e^1\end{pmatrix}
-2
\begin{pmatrix}
 2 b^3  \cr
 1 + (b^3)^2 - \left(\rme^{h} +\ft12E \right)^2  \cr
- b^3 e^1 + e^2 \left(\rme^{h} +\ft12E \right)\cr
 b^3 e^2 - e^1 \left(\rme^{h} +\ft12E \right)
\end{pmatrix},
  \label{kSU2}
\end{equation}
where $E\equiv (e^1)^2 + (e^2)^2$. The moment map is then
\begin{equation}
  {\cal P}=
  \xi \begin{pmatrix}
  -\sqrt{2}\rme^{-h/2}e^2\cr
   \sqrt{2}\rme^{-h/2}e^1\cr
   1-\ft12\rme^{-h}E\end{pmatrix} + 
   \begin{pmatrix}
   -\sqrt{2}\rme^{-h/2}\left[b^3e^1+e^2 \left(-\rme^{h}+\ft12E\right)\right]  \cr
  \sqrt{2}\rme^{-h/2}\left[-b^3e^2+e^1 \left(-\rme^{h}+\ft12E\right)\right]  \cr
  -\ft12\rme^h-\ft12\rme^{-h}\left[ 1+(b^3)^2+\ft14E^2\right] +\ft32E
   \end{pmatrix}
 \label{PSU2}
\end{equation}
After truncation (\ref{e1e20}), we get the following $D$-term
\begin{equation}
D = -g\left[(b^3)^2\rme^{-h}+\rme^{-h}+ \rme^{h}\right] +2g \xi.
\end{equation}

\subsection{Common formulae in the truncated space}
 \label{ss:commonformulae}
The data of the two models lead to common formulae in the truncated
space. The quaternionic vielbein as obtained from
appendix~\ref{app:paramCosets}, and with the conditions (\ref{b1b20}) or
(\ref{e1e20}), reduces to
\begin{equation}
f^{11}=- \rmi\alpha  \frac{\rmd\Phi }{2\Im\Phi },\qquad
 f^{22}=\rmi\alpha \frac{\rmd\bar \Phi }{2\Im\Phi },\qquad
 f^{12}=f^{21}=0,
\end{equation}
where
\begin{equation}
  \alpha =\sqrt{2} \quad \mbox{for }\frac{\rSp(1,1)}{\rSp(1)\rSp(1)},
  \qquad \alpha =1 \quad \mbox{for }\frac{\rSU(2,1)}{\rSU(2)\U(1)}.
 \label{constantnormalization}
\end{equation}
In terms of the complex field $\Phi$, the metric is
\begin{equation}
  \rmd s^2=g_{XY}\rmd q^X\rmd q^Y=\frac{\alpha ^2}{2(\Im\Phi)^2 }\rmd\Phi
  \rmd\bar \Phi .
 \label{ds2common}
\end{equation}
The Killing vector $k$ reduces to
\begin{equation}\label{red.killing}
k= -2(\Phi^2+1)\partial_\Phi + {\rm c.c.}
\end{equation}
The $D$-term in terms of $\Phi$ is given by
\begin{equation}
D\equiv 2g {\cal P}^3=-g\alpha ^2 \frac{|\Phi |^2+1}{\Im\Phi }+2g\xi.
 \label{P3general}
\end{equation}
Note that  both our examples give the same $D$-term, up to a
normalization in the first term. The normalization issue is not related
to remaining fields in this reduction, but is due to the non-Abelian
aspects of the R-symmetry $\SU(2)$ in the different quaternionic-K{\"a}hler
manifolds, leading to a different normalization for
\begin{equation}
  \omega ^3= \alpha ^2\frac{\rmd\Phi +\rmd\bar \Phi }{8\Im\Phi }.
 \label{omega3Phi}
\end{equation}

The complex field $\Phi$ belongs to a chiral multiplet in the resulting
$N=1$ theory and parametrizes the \emph{non-trivial} K{\"a}hler space
$\rSU(1,1)/\U(1)$. Further on, we will show that the $N=2$ BPS equations
reduce correctly to those of this $N=1$ system.

\section{The $N=2$ BPS equations}
\label{ss:TruncN2BPS}

We study the $N=2$ BPS equations for a system with one vector- and one
hypermultiplet, and the truncation to $N=1$. The hypermultiplet target
space and gauging were described above. Below, we define the special
geometry of the vector multiplet target space.

\subsection{Choice of special geometry}

The special geometry that we consider is  the  {\em minimal} one defined by the quadratic prepotential
\begin{equation}
F(X^0,  X^1)=-\ft12\rmi\left[{X^0 X^0-X^1 X^1}\right]
\label{prepotential}.
\end{equation}
This leads to $N_{IJ}=2\eta _{IJ}$ with $\eta _{IJ}={\mathop{\rm
diag}\nolimits}(-1,1)$. Introducing the special coordinate
$z=\frac{X^1}{X^0}$, we obtain the K{\"a}hler potential (\ref{NdefResult})
\begin{equation}
\cK=-\log\left[{2(1-z\bar z)}\right],\qquad g_{z\bar z}=(1-z\bar z)^{-2},
 \label{cKmodel}
\end{equation}
where we replace the only value of the index $\alpha $ with $z$. This
corresponds to the coset $\frac{\rSU(1,1)}{\rU(1)}$. We also obtain the
vector kinetic matrix
\begin{equation}
\cN_{IJ} =-\frac{\ii}{1-z^2}
\begin{pmatrix}
1+z^2 & -2z \\
-2 z & 1+z^2
\end{pmatrix}.
\end{equation}
At the base point $z=0$, which we will use in the reduction to $N=1$, we
have thus $\Im{\cal N}_{IJ}=-\delta _{IJ}$ and $\Re{\cal N}_{IJ}=0$.

The symplectic section obtained from the prepotential according to
(\ref{Vlocal}) is
\begin{equation} V =\begin{pmatrix}
X^0 \\
X^1 \\
-\ii X^0\\
\ii X^1
\end{pmatrix}= \rme ^{{\cal K}/2}
\begin{pmatrix}
1 \\
z \\
-\ii \\
\ii z
\end{pmatrix}, \qquad
{\cal D}_zV= \frac{\rme^{{\cal K}/2}}{1-z \bar z}\begin{pmatrix} \bar z \\
1
\\ -\ii \bar z \\ \ii \end{pmatrix}. \label{symplsectionsSU2}
\end{equation}

We intend to use the vector $W_\mu\equiv W^1_\mu$ to gauge the
appropriate isometry with the Killing vector $k$. To compute the scalar
potential we need the component $U^{11}$ in (\ref{formulaU}) which is
easily computed using (\ref{symplsectionsSU2}) such that
\begin{equation}
U^{11}-3X^1\bar X^1=\rme^{{\cal K}}(1-3 z\bar z)= \frac{1-3 z\bar
z}{2(1-z \bar z)}.
\end{equation}
The scalar potential is then given by
\begin{equation}
g^{-2}\cV  = \frac{2(1-3 z \bar z)}{1- z\bar z}\cP^x\cP^x+\frac{z\bar
z}{(1- z\bar z)} g_{XY} k^X k^Y , \label{potential}
\end{equation}
where $\cP^x$ is the moment map that corresponds to the Killing vector $k$.

\subsection{Ansatz for the bosonic fields in $N=2$}

Motivated by the fact that we want our $N=2$ bosonic action to reduce to
the one of an $N=1$ theory, we look for a field configuration that
effectively truncates the $N=2$ action with vector- and hypermultiplet to
an $N=1$ action with vector (i.e. gauge) and chiral multiplet. Consistent
truncations of $N=2$ to $N=1$ are studied in \cite{Andrianopoli:2001zh},
to which we refer for details. The consistency conditions derived there
come from demanding:
\begin{equation}
\delta \psi_{2\mu} = 0 \quad \mbox{with} \quad \epsilon_2 = 0 ,
 \label{eps20}
\end{equation}
(similarly for the other truncated fermions). This choice is consistent
with the survival of the complex structure $J^3$ in the reduced theory.
Furthermore, we demand that the sources of truncated bosonic fields
vanish. The conditions can be satisfied by imposing the following
conditions on the bosonic field configuration:
\begin{enumerate}
\item  The scalar $z$ of the vector multiplet vanishes on the configuration.
\item  We gauge a $\U(1)$ isometry of the quaternionic-K{\"a}hler manifold with the vector field $W\equiv W^1$.
The  gauging is Abelian (therefore the symmetries of the special manifold
are not gauged) and the bare graviphoton does not gauge any symmetries
and is put to zero: $W^0=0$.
\item The only non-vanishing components of the moment map $\cP^x$  and the
quaternionic $\rSU(2)$ connections $\omega^x$ are $\cP^3$ and $\omega^3$,
respectively.
 \end{enumerate}
These conditions are implemented  as follows:
\begin{eqnarray}
z =W^0 &=& 0,\nonumber\\
 k_0^X =
  k^z _I&=& 0,\nonumber\\
\cP^1   = \cP^2 &=& 0, \nonumber\\
 \omega^1 = \omega^2 & =& 0.
 \label{condition1}
\end{eqnarray}
Note that $z=0$ is a critical point of the scalar potential
(\ref{potential}).

With our choice of special geometry based on the quadratic prepotential
(\ref{prepotential}), the conditions above imply that on the
configuration
\begin{equation}
g_{z \bar z }=1,\quad  {\cal D}_z X^I =\frac{1}{\sqrt{2}}
 \left(\begin{array}{c} 0 \\ 1 \end{array}\right) , \quad \cN_{IJ}=-\rmi\left(\begin{array}{cc} 1 & 0 \\ 0 & 1\end{array}\right).
 \label{reduceddata}
 \end{equation}
This implies:
\begin{equation}
S^{ij}={\cal N}^{iA}=T^-_{\mu\nu} =A_\mu =0.
 \label{condition2}
\end{equation}
The non-vanishing data are
\begin{equation}
V_i{}^j=\rmi\left(\omega ^3 + g W {\cal P}^3\right)
\left(\sigma^3\right)_i{}^j,\qquad N^z_{ij}=\frac{1}{\sqrt{2}}\rmi{\cal
P}^3\left(\sigma^3\right)_{ij},
 \label{nonvanishingdata}
\end{equation}
where
\begin{equation}
(\sigma^3)_i{}^j = \left(\begin{array}{cc} 1 & 0\\ 0 &
-1\end{array}\right)   , \qquad (\sigma^3)_{ij}=\left(\begin{array}{cc} 0
& 1\\ 1 & 0  \end{array}\right).
\end{equation}
The data of the quaternionic manifold are in
section~\ref{ss:commonformulae}.

On the configuration defined above, the bosonic part of the $N=2$ action
can be obtained from (\ref{L4leading}) minus the potential of
(\ref{potential}). This gives
\begin{equation}
e^{-1}\cL= \frac{1}{2} R  -\frac{1}{4}F^{\mu\nu}F_{\mu\nu} -\frac{\alpha
^2}{4(\Im \Phi)^2} \nabla  _\mu \Phi\nabla  ^\mu \bar \Phi-2g^2
\left[\alpha ^2 { \frac{\abs{\Phi}^2+1}{2\Im \Phi}-\xi }\, \right]^2,
\end{equation}
where
\begin{equation}
  \nabla _\mu\Phi =\partial _\mu \Phi-2gW_\mu \left( \Phi ^2+1\right).
 \label{nablaPhi}
\end{equation}
Note that since the ansatz satisfies the conditions for a consistent
truncation, solutions of the field equations derived from this action are
solutions of the full $N=2$ field equations. This is due to the fact that
the truncated fields appear at least quadratically in the $N=2$ action.

The supersymmetry transformations (\ref{N=2susyeqns}) become, using
(\ref{condition1}) and (\ref{condition2}),
\begin{eqnarray}
 \delta \psi _\mu ^i & =  & \nabla _\mu(\omega ) \epsilon^i
 =\left( \partial_\mu + \ft14 \omega_\mu{}^{ab} \gamma _{ab}\right) \epsilon ^i
 + V_{\mu j}{}^i\epsilon ^j,\nonumber\\
 \delta \lambda ^z_ i&=& -\ft12g^{z\bar z }{\cal D}_{\bar z}
 \bar X^1\Im{\cal N}_{11}F^{-1}_{\mu \nu }\gamma ^{\mu \nu }\varepsilon _{ij}\epsilon^j
+ gN^z _{ij}\epsilon ^j,\nonumber\\
 \delta \zeta ^A&=&\ft12\rmi f^{Ai}_X\slashed{\nabla }q^X\epsilon_i.
\end{eqnarray}
Using (\ref{eps20}), (\ref{reduceddata}) and (\ref{nonvanishingdata}),
this gives us
\begin{align}
\delta \psi_{\mu}^1 &=  (\partial_\mu +\ft{1}{4}\omega_
{\mu|ab}\gamma^{ab}+ \ft12\ii A^B_\mu )\,\epsilon^1,  & \delta
\psi_{\mu}^2 &=  (\partial_\mu +\ft{1}{4}\omega_
{\mu|ab}\gamma^{ab}-\ft12\ii A^B_\mu )\,\epsilon^2,\nonumber\\
 \delta\lambda_2 & =
 - \frac{1}{2\sqrt{2}}F^{-}_{\mu\nu}\gamma^{\mu\nu}\epsilon^1
-\ii  \frac{1}{\sqrt{2}}D\, \epsilon^1,
  &
\delta \lambda_1 & =  \frac{1}{2\sqrt{2}}
F^{-}_{\mu\nu}\gamma^{\mu\nu}\epsilon^2 -\ii \frac{1}{\sqrt{2}} D\,
\epsilon^2,
\nonumber\\
 \delta \zeta^1 &=  \frac{\alpha }{4\Im\, \Phi}\slashed{\nabla} \Phi \epsilon_1,   &
 \delta \zeta^2 &= - \frac{\alpha }{4\Im\, \Phi}\slashed{\nabla}\bar \Phi \epsilon_2.   \label{N=1BPS}
\end{align}
In these equations  $A^B_\mu $ is the  matter connection of the gravitini
on the configuration:\footnote{In general for such reductions to $N=1$
using only $\epsilon ^1$, we have $A_\mu ^B=A_\mu -2\rmi V_{\mu 1}{}^1$.}
\begin{equation}
A^B_\mu=2\omega^3_\mu +W_\mu D= \frac{\alpha ^2\left( \partial _\mu \Phi
+\partial _\mu\bar \Phi\right)  }{4\Im\Phi }+W_\mu D.
 \label{ABfromN2}
\end{equation}
We see that the equations for $\epsilon_1$ and $\epsilon_2$ split into
two sets. Starting from the $N=2$ action, we have defined a consistent
truncation $N=2 \rightarrow N=1$: once we take $\epsilon_2\equiv0$, the
BPS equations for $\epsilon_1$, $\delta \psi _\mu ^1=\delta \lambda
_2=\delta \zeta ^1=0$, correspond to the BPS equations of an $N=1$
supergravity theory with vanishing superpotential, a $D$-term, a constant
effective coupling for the vector field kinetic term, and a $\U(1)$
gauging of the isometry $\delta \Phi = 2g(\Phi^2+1)$ of the upper half
plane. We verify this explicitly in section~\ref{sss:N=1BPS}.

However, if the equations for $\epsilon_2$, $\delta \psi _\mu ^2=\delta
\lambda _1=\delta \zeta ^2=0$, can be simultaneously solved with those
for $\epsilon_1$, we have an $N=2$ BPS solution with an extra
supersymmetry corresponding to $\epsilon_2$. In the next section we will
show this to be the case for a (1/2)-projection of both supersymmetries.
{}From now on we will continue with both sets of BPS equations (for
$\epsilon_1$ and $\epsilon_2$) of the full N=2 theory.

For similar results on such an extra supersymmetry arising from an
embedding of $N=1$ into $N=2$ in the context of global supersymmetry, see
\cite{Achucarro:2004ry} and \cite{Achucarro:2002jg}.

\section{Finding a BPS cosmic string solution}
\label{ss:resultN1}

We now proceed to solve the resulting BPS equations for a cosmic string ansatz.
We study the BPS equations for $\epsilon_1$ and $\epsilon_2$ together.

\subsection{String ansatz; projector and integrability conditions}

We assume a straight, static cosmic string on the $z$-axis. We use
cylindrical coordinates $(t,z,r,\theta )$.  We take the following ansatz
for the spacetime metric:
\begin{equation}
  \rmd s^2 = -\rmd
t^2+\rmd z^2+\rmd r^2+C^2(r)\rmd\theta^2. \label{ansatzmetric}
\end{equation}
The vielbeins are (we take $C(r)>0$ without loss of generality)
\begin{equation}\label{spacetimeVielbein}
\hat e^1 = \rmd r, \qquad \hat e^2= C(r)\rmd\theta,
\end{equation}
from which we can deduce the spin connection
\begin{equation}\label{stconnection}
\omega_ r^{12} = 0, \qquad \omega_ \theta^{12}=-C'(r)\equiv -\frac{\rmd\,
C(r)}{\rmd\, r}.
\end{equation}
The complex field is independent of $z$: $\Phi=\Phi(r,\theta)$.

Squaring the BPS equations for the chiral fermions, one gets a
consistency condition for the projectors on the Killing spinors. The
projector condition can also be derived from the integrability conditions
\begin{equation}
 (C''\gamma^{12} -\ii  F^B_{r\theta})\epsilon^1 = 0  , \qquad
( C''\gamma^{12} + \ii  F^B_{r\theta})\epsilon^2 = 0.
 \label{integrability2}
\end{equation}
One obtains in this way that
\begin{equation} \label{bpsproj}
\gamma^{12}\epsilon^1 =\mp \ii\epsilon^1, \qquad  \gamma^{12}\epsilon^2
=\pm \ii\epsilon^2.
\end{equation}
(so $\epsilon ^1$ and $\epsilon ^2$ have opposite chirality on the string
worldsheet). It follows that the string configuration preserves a maximum
of 4 supercharges out of the 8 supercharges of the $N=2$ supergravity
system. Imposing this projector condition gives the following BPS
equations, which follow from (\ref{N=1BPS}):
\begin{eqnarray}
(\partial_\mu \mp  \frac{\ii}{2}\omega_ {\mu|12}+ \frac{\ii}{2}
A^B_\mu ) \epsilon^1& =&  0,  \label{gravitinibps1}\\
(\partial_\mu \pm  \frac{\rmi}{2}\omega_ {\mu|12}-\frac{\ii}{2}A^B_\mu
)\,\epsilon^2  &=&  0,  \label{gravitinibps2}
\\
\mp  C^{-1}   F_{r\theta}+\, D  & = & 0  ,\label{gauginibps} \\
(\nabla_r \pm \ii C^{-1}\nabla_\theta)\Phi &=& 0,  \label{hyperinibps}
\end{eqnarray}
and the integrability condition
\begin{equation}
C''\pm F_{r\theta }^B=0.
\end{equation}
\subsection{Solving the BPS equations}
To find a half BPS cosmic string solution of the $N=2$ theory, we attempt to solve the BPS equations given above.

\subsubsection*{Hyperini BPS equations and ansatz for the scalar}

To solve the  hyperini BPS equation (\ref{hyperinibps}), we  will follow
\cite{Taubes:1979tm} by  defining a  holomorphic derivative on the  plane
perpendicular to the cosmic string. This is possible because any two
dimensional metric is K{\"a}hler and therefore admits a complex structure.
This property  has been used to obtain BPS equations for cosmic strings
by Comtet and Gibbons \cite{Comtet:1988wi}, see also Ruback
\cite{Ruback:1987sg}. The use of holomorphic derivatives will give us a
nice way to get the right ansatz for the scalar field. The method will be
seen to amount to a coordinate transformation of the upper half plane to
the unit disk. Let us define
\begin{equation}
  z=\exp\left[ \int \frac{\rmd r}{C(r)}+ \rmi\theta\right]  .
 \label{defzrtheta}
\end{equation}
With these coordinates the 2-dimensional metric is $\rmd s^2=\Omega
^2\rmd z \rmd\bar z$, where $\Omega $ is the conformal factor that
reduces to 1 when $z=0$. We then have
\begin{equation}
z\partial_ z = C \partial_r\mp \ii \partial_\theta,\qquad \bar
z\partial_{ \bar z}= C \partial_r\pm \ii \partial_\theta,
\end{equation}
and
\begin{equation}
zW_z= C W_r\mp \ii W_\theta, \qquad \bar z W_{\bar z}= C W_r\pm \ii
W_\theta.
\end{equation}
We can then write the hyperini BPS equation (\ref{hyperinibps}) as
\begin{equation}
\nabla _{\bar z}\Phi =
\partial_{\bar z}\Phi+gk^{\Phi} W_{\bar z}=0, \qquad k^{\Phi }=-2\left( \Phi
^2+1\right) .
\end{equation}
where  $\delta \Phi=-gk^\Phi$, in our case $\delta \Phi=2g(\Phi^2+1)$. We
can now  solve the equation for $W_z$:
\begin{equation}
2gW_{\bar z}=\frac{\partial_{\bar z} \Phi}{\Phi^2+1}=\partial_{\bar z}
\tan^{-1}(\Phi)
\end{equation}
Using the identity
\begin{equation}
\tan^{-1} \Phi= \frac{\ii}{2}\log\frac{\ii+\Phi}{\ii-\Phi},
\end{equation}
we have
\begin{equation} \label{Az}
2gW_{\bar z}=-\frac{\ii}{2}\partial_{\bar z} \log u,
\end{equation}
where we define
\begin{equation} \label{cotrafo}
u= \frac{\ii-\Phi}{\ii+\Phi},\qquad \mbox{ that is  } \qquad
\Phi=\ii\frac{1-u}{1+u}.
\end{equation}
The expression (\ref{Az}) is familiar from the study of Abelian vortices
in flat space. The variable $u$ is convenient to analyse the BPS
equations, as we see by looking at the gauge transformation and the
$D$-term.  In the $u$-plane, the gauge transformation is a change of
phase with charge $+4g$:
\begin{equation}
\delta u= 4g\ii u ,
\end{equation}
and the $D$-term is a function of $\abs{u}^ 2$:
\begin{eqnarray}
D =-2 g\alpha ^2\frac{1+ \abs{u}^2}{1-\abs{u}^2}+2g\xi.
\end{eqnarray}
If $\xi > \alpha^2$, the moduli of vacua in the $u$-plane is a circle
centered at the origin $(u=0)$ with its  radius  fixed by the FI term
$\xi$:
\begin{equation}
D=0\iff \abs{u}^2=\frac{\xi-\alpha ^2}{\xi+\alpha ^2}.
\end{equation}
Note that, if $\xi = \alpha^2$, there is a unique vacuum $u=0$ and the
$\U(1)$ gauge symmetry is not spontaneously broken. For $\xi < \alpha^2$
the vacuum is an unstable de Sitter solution. In what follows we assume
$\xi > \alpha^2$.

For a cosmic string located at the origin, we now take the following
ansatz:
\begin{equation}
u=f(r)\rme^{\ii n\theta}, \quad f(0)=0 , \quad f(\infty)\rightarrow
\sqrt{\frac{\xi-\alpha ^2}{\xi+\alpha ^2}}.
\end{equation}
Using the identity
\begin{equation}
\log u = \ft{1}{2}\log\abs{u}^2 +\ii(\arg u+2\pi m),
\end{equation}
it follows that
\begin{equation}
2gW_{\bar z}=-\ft12{\ii}\partial_{\bar z} \log  u=-\ft12{\ii}
(\ft{1}{2}C(r)\partial_r \log \abs{u}^2\mp C^{-1} n),
\end{equation}
or in other words
\begin{equation}
W_r=0, \quad \mp 2gW_\theta= \frac{1}{2}\left[{C(r) \frac{f'(r)}{f(r)}\mp
n}\right]. \label{profile1}
\end{equation}

\subsubsection*{Gaugini BPS equation}

We can now solve the BPS equations of the gaugini. First we need to
compute the field strength. As we work in the gauge $W_r=0$, we have
\begin{equation}
F_{r\theta}= \partial_r W_\theta ,
\end{equation}
and the  gaugini BPS equation is
\begin{equation}
\pm  W_\theta'(r)=C(r) \,D. \label{profile2}
\end{equation}
\subsubsection*{Gravitini BPS equations}

The gravitini BPS equations are
\begin{align}
(\partial_r  +\frac{\ii}{2}A_r^B) \epsilon^1&=0,   & [\partial_\theta \pm \frac{\ii}{2}C'(r)+\frac{\ii}{2}A^B_\theta] \epsilon^1 &=0, \\
(\partial_r  -\frac{\ii}{2}A_r^B )\epsilon^2 &=0,   & [\partial_\theta
\mp \frac{\ii}{2}C'(r)-\frac{\ii}{2}A^B_\theta] \epsilon^2 &=0,
\end{align}
with the integrability condition
\begin{equation}
 C''\pm F^B_{r\theta }=0.
\end{equation}
These equations are different from those solved in \cite{Dvali:2003zh},
because here the radial component $A_r^B$ of the gravitini connection
$A^B_\mu$ does not vanish. This can be traced back to the non-vanishing
of the radial component of the K{\"a}hler connection of the half-plane. The
complex scalar parametrizes the K{\"a}hler-Hodge manifold
$\frac{\rSU(1,1)}{\rU(1)}$ with the upper half plane K{\"a}hler potential:
 \begin{equation}
 \cK^{\rm HP}= -\alpha^2\log -\ii(\Phi-\overline\Phi).
\end{equation}
Under the change of variables (\ref{cotrafo}) it gets the form
\begin{equation}
\cK^{\rm HP}=-\alpha^2\log \frac{2(1-u\bar u)}{(1+u)(1+\bar
u)}=-\alpha^2\log 2(1-u\bar u)+\alpha^2\log (1+u)+\alpha^2\log(1+{\bar
u}).
\end{equation}
The K{\"a}hler potential $\cK^{\rm HP}$  is not invariant under the symmetry
$\delta \Phi =2g\left( \Phi^2+1\right) $, but the manifold
$\frac{\rSU(1,1)}{\rU(1)}$ admits another K{\"a}hler description in term of
the  unit disk  K{\"a}hler potential
\begin{equation}
\cK^{\rm UD}=-\alpha^2\log 2(1-u\bar u),
\end{equation}
which is invariant under the symmetry $\delta u =4g\ii u$. As we have
\begin{equation}
\cK^{\rm HP}= \cK^{\rm UD} +\ell+\bar\ell\quad \text{ with } \quad
\ell(u)=\alpha^2\log (1+u),
 \label{cKHPell}
\end{equation}
we see that the upper half plane and the unit disk are related by a
K{\"a}hler transformation generated by the analytic function $\ell(u)$. As is
well known, the K{\"a}hler metric does not change under a K{\"a}hler
transformation, however it is  the K{\"a}hler $\rU(1)$-connection that enters
the BPS equation of the gravitino
\begin{equation}
\cQ= -\frac{\ii}{2}(\rmd \phi \frac{\partial\cK}{\partial \phi}-\rmd
\overline{\phi} \frac{\partial\cK}{\partial\overline{\phi}}), \qquad A^B
= \cQ +W D,
\end{equation}
and the latter  transforms as
\begin{equation}
\cQ\rightarrow \cQ -\frac{\ii}{2}(\rmd \phi \frac{\partial \ell}{\partial \phi}-\rmd \bar\phi \frac{\partial \bar\ell }{\partial\overline{\phi}})
=\cQ +\rmd \Im \ell.
\end{equation}
{}From the point of view of $N=2$ supergravity, the K{\"a}hler potential
comes from the metric of the quaternionic-K{\"a}hler manifold, and the choice
of the half-plane was imposed by the value of the quaternionic
$\rSU(2)$-connection $\omega^x$ which becomes the $\rU(1)$-connection of
the K{\"a}hler-Hodge manifold defined by the string configuration.

Clearly, the K{\"a}hler transformation of the K{\"a}hler-Hodge manifold is
related to a change of gauge of the $\rSU(2)$-connection $\omega^x$,
which is in the normalization of \cite{Bergshoeff:2004nf}
\begin{equation}
\omega^x \rightarrow \omega^x-\ft12\nabla \ell ^x.
\end{equation}
As we have $\omega^1=\omega^2=0$ on the string configuration, we see that
if  $r^1=r^2=0$ on the string configuration, the $\rSU(2)$-redefinition
of the quaternionic connection is seen as a K{\"a}hler transformation for the
K{\"a}hler-Hodge manifold defined by the string configuration. Comparing with
(\ref{ABfromN2}), we have
\begin{equation}
\Im\ell =-\ell ^3.
\end{equation}
Using the analytic property of $\ell$, the transformation of the $\rU(1)$ connection under the K{\"a}hler transformation generated by $\ell$ can be rewritten as:
\begin{equation}
\cQ \rightarrow \cQ-\rmd \left({\,
\ii{\frac{\ell-\bar{\ell}}{2}}}\right),
\end{equation}
so that we have
\begin{equation}
\cQ^{\rm HP} = \cQ^{\rm UD}-\ii
\rmd\left({\frac{\ell-\overline{\ell}}{2}}\right),
\end{equation}
where $\cQ^{\rm UD}$ is the  $\rU(1)$-connection of the unit disk:
\begin{equation}
\cQ^{\rm UD} =\ii\frac{\alpha ^2}{2}\frac{u\rmd\overline{u}-\rmd
u\overline{u}}{1-u\overline{u}}.
\end{equation}
To solve the BPS equation we use $u=f(r)\rme^{\ii n\theta}$. This implies for the unit disk:
\begin{equation}
\cQ^{\rm UD}_r = 0, \qquad  \cQ^{\rm UD}_\theta=n\alpha
^2\frac{f^2}{1-f^2},
\end{equation}
whereas the half-plane gets an extra contribution coming from $\ell$,
which introduces a dependence on the azimuthal angle. As $W_r=0$, $A^B$
is just a function of the radius $r$ in the case of the unit disk and the
gravitini BPS equations give differential equations for the profile
functions  $f(r)$ and $C(r)$ depending only on one variable $r$.

The situation is not that nice for the half-plane, due to the presence of
$\ell$, see (\ref{cKHPell}), which depends explicitly on the azimuthal
angle $\theta$. It is interesting to note that  we could actually work
with the unit disk K{\"a}hler potential if we had defined our quaternionic
structure with the   $\rSU(2)$-connection $\omega^x+\ft12\rmd(\Im \ell)$.

To solve the BPS equations for the half-plane, we redefine the Killing
spinors by a rotation, in order to make the gravitini BPS equations
independent of the azimuthal angle:
\begin{equation}
\epsilon^1 = \exp \left({\frac{\ii}{2}\Im\,\ell}\right)\tilde\epsilon^1 ,
\qquad \epsilon^2 = \exp
\left(-{\frac{\ii}{2}\Im\,\ell}\right)\tilde\epsilon^2
\end{equation}
Such a redefinition has the same effect as a K{\"a}hler transformation.
Indeed, the gravitini BPS equations for $\tilde{\epsilon^i}$ are
\begin{equation}
 \left[{\cD +\ft12\ii\rmd (\Im\,\ell)+\ft12\ii A^B}\right] \tilde\epsilon^1=0, \qquad
 \left[{\cD -\ft12\ii\rmd(\Im\,\ell )-\ft12\ii A^B}\right] \tilde\epsilon^2=0,
\end{equation}
where ${\cal D}$ is a derivative including spin connection. Defining
\begin{equation}
\tilde A^B= A^B+\rmd (\Im \, \ell),
\end{equation}
we have
\begin{equation}
\tilde A^B =\cQ^{\rm UD}+W\,D.
\end{equation}
In terms of   $\tilde\epsilon^i$ and $\tilde A^B$, the gravitini BPS
equations  have the same structure as  in  \cite{Dvali:2003zh}:
\begin{align}
\partial_r  \tilde\epsilon^1&=0   & [\partial_\theta \pm \ft12{\ii}C'(r)+\ft12{\ii}\tilde A^B_\theta] \tilde\epsilon^1 &=0, \label{gravitinibpsfinal1}\\
\partial_r \tilde\epsilon^2 &=0   & [\partial_\theta \mp \ft12{\ii}C'(r)-\ft12{\ii}\tilde A^B_\theta] \tilde\epsilon^2 &=0, \label{gravitinibpsfinal2}
\end{align}
where $\tilde A^B_\theta$ depends only on the radial distance $r$.

We can now follow the treatment of  \cite{Dvali:2003zh} to solve the BPS equations.
Globally well defined spinors are
\begin{equation}
\tilde\epsilon^1 = \rme^{\mp \frac{\ii}{2}\theta}\tilde\epsilon_0^1 \iff
\tilde\epsilon_2= \rme^{\pm \frac{\ii}{2}\theta}\tilde\epsilon_0^2,
\end{equation}
where $\epsilon_0^i$ is a constant spinor that satisfies the same
projection relation as $\epsilon^i$.

Therefore, the  gravitini BPS equations are equivalent to the differential equation
\begin{equation}
C'= 1\mp \tilde A^B_\theta. \label{profile3}
\end{equation}
and the Killing spinors   $\epsilon^i$ are related to $\tilde\epsilon^i$
by a non-constant shift of phase:
\begin{equation}
\epsilon^1 = \rme^{\ii\Delta(r,\theta)}\,\,\tilde\epsilon^1,\quad
\epsilon^2 = \rme^{-\ii\Delta(r,\theta)}\,\,\tilde\epsilon^2,
\end{equation}
where
\begin{align}
\Delta(r,\theta) &=\frac{1}{2}\Im\,\ell=-\frac{\ii\alpha ^2}{4}\log \frac{1+u}{1+\overline{u}}\\
& =\frac{\alpha^2}{2} \arg (1+u)=\frac{\alpha^2}{2}\tan^{-1}
\left({\frac{f(r) \sin n\theta}{1+f(r)\cos n\theta}}\right).
\end{align}
When  $r\rightarrow \infty$, $\Delta(r,\theta)$  does not depend on $r$
anymore, as $f\rightarrow \sqrt{\frac{\xi-\alpha ^2}{\xi+\alpha ^2}}$.

\subsection{Profile of the string}

{}From equations (\ref{profile1}), (\ref{profile2}) and (\ref{profile3})
we obtain  the equations that determine the profile of the string:
\begin{eqnarray}
\pm f'(r) &=& \frac{f(r)}{C(r)}\left({n-4gW_{\theta}(r)}\right), \nonumber\\
\pm W'_{\theta}(r)&=&C(r)D(r),\nonumber\\
 C'(r)&=&1\mp \tilde A^B_{\theta}(r),
\end{eqnarray}
with
\begin{eqnarray}
\tilde A^B_{\theta} &=&\cQ^{\rm UD}_{\theta}+W_{\theta} D,\nonumber\\
\cQ^{\rm UD}_\theta&=&n\alpha ^2\frac{f^2}{1-f^2},\nonumber\\
D&=&-2g\alpha ^2\frac{1+f^2}{1-f^2}+2g\xi,
\end{eqnarray}
and asymptotic behaviour
\begin{eqnarray}
f &\sim& \mbox{ const } r^{\pm n}, \quad C \sim r, \quad W_{\theta} \sim
\pm g(\xi - \alpha ^2) r^2 \ \mbox{ for } \ r \rightarrow 0, \nonumber\\
 f
&\rightarrow& \sqrt{\frac{\xi-\alpha ^2}{\xi+\alpha ^2}}, \quad
W_{\theta} \rightarrow \frac {n}{4g}
 \ \mbox{ for } \ r \rightarrow \infty.
\end{eqnarray}
The upper or lower sign apply for positive or negative winding number
$n$, respectively.

The metric for $r\rightarrow \infty$ is given by
\begin{equation}
\rmd s^2 = -\rmd t^2+\rmd z^2+\rmd r^2+r^2\left[1\mp \ft12n(\xi-\alpha
^2)\right]^2\rmd\theta^2.
\end{equation}

The asymptotic behaviour is similar to the case of \cite{Dvali:2003zh}.
At $r\rightarrow \infty$, the string creates a locally-flat conical
metric with a deficit angle proportional to $\xi -\alpha ^2$.  The energy
of the string per unit length can be computed as in \cite{Dvali:2003zh}.
The details of the calculation are given in section \ref{sss:BogKah}
below: one finds that  the only non-vanishing contribution comes from the
Gibbons-Hawking surface term \cite{Gibbons:1976ue}
\begin{equation}
\mu_{\mbox{string}} = -\left.\int \rmd\theta\, C'\right|_{r=\infty } +
\left.\int \rmd\theta\, C'\right|_{r=0} =\pm  \pi n(\xi-\alpha ^2)>0.
\end{equation}
Note also that the full $N=2$ supersymmetry is restored asymptotically.

\section{$N=1$ $D$-term strings with arbitrary K{\"a}hler potentials}
\label{s:N=1Sugra}

\subsection{Comparison with $N=1$ supergravity}
\label{sss:N=1BPS}

First we check that the BPS equations derived from the truncated $N=2$
theory are consistent with the general expressions for $N=1$ supergravity
\cite{Cremmer:1983en}, which in the present form can be found in
\cite{Kallosh:2000ve,Binetruy:2004hh}. The $N=1$ action, completely
determined by the K{\"a}hler-potential ${\cal K}(\phi,\phi^*)$, the
holomorphic function $f_{\alpha \beta}(\phi)$ (no superpotential), the
gauging and the FI terms, is\footnote{Note that for easy comparison with
the $N=1$ papers, the index $i$ now refers to chiral multiplets, and thus
will only take one value in our example: $\phi _i=\phi$ and $\phi ^i=\bar
\phi$. For the fermions: $\chi _i=\chi _L$ and $\chi ^i=\chi _R$. On the
other hand $\alpha $ now refers to the different vector multiplets.}
\begin{equation}
e^{-1}{\cal L}_{\rm bos}=\ft12 R -g_i{}^j(\nabla_\mu \phi ^i)(\nabla^\mu
\phi _j) -V_D -\ft14(\Re f_{\alpha \beta}) F_{\mu \nu }^\alpha F^{\mu \nu
\,\beta }
 +\ft 18 e^{-1}\varepsilon ^{\mu \nu \rho \sigma }(\Im f_{\alpha \beta})
 F_{\mu \nu }^\alpha F_{\rho \sigma }^{\beta }.
 \label{bosonic action}
\end{equation}
with covariant derivative given by
\begin{eqnarray}
\nabla_\mu \phi ^i = \partial_\mu \phi ^i + gk_\alpha^i\, W_\mu^\alpha.
\end{eqnarray}
The potential consists only of a $D$-term:
\begin{eqnarray}
  V_D&=&\ft12(\Re f_{\alpha \beta}) D^\alpha D^\beta=
 \ft{1}{2}(\Re f)^{-1\,\alpha \beta }  {\cal P}_\alpha {\cal P}_\beta,
\label{V_D}
\end{eqnarray}
where
\begin{equation}
\partial _i{\cal P}_\alpha(\phi,\phi^*)=-\rmi g\,
k_{\alpha j}g^j{}_i. \label{defPN1}
\end{equation}
Here, the moment map appears only differentiated, and one can thus add an
arbitrary constant, which is the FI term.

The $N=1$ supersymmetry transformations for a bosonic configuration are given by:
\begin{eqnarray}
\delta \psi _{\mu L}  & = &
\left( \partial _\mu  +\ft14 \omega _\mu {}^{ab}(e)\gamma _{ab}
+\ft 12\rmi A_\mu^B \right)\epsilon_L 
, \nonumber\\
\delta \chi _i& = & \ft12\not\! \nabla\phi_i \epsilon_R,
 \nonumber\\
\delta \lambda^\alpha  &=&\ft14\gamma ^{\mu \nu } F_{\mu \nu
}^\alpha\epsilon +\ft12\rmi \gamma _5 (\Re f)^{-1\,\alpha \beta}{\cal P}
_\beta \epsilon  \label{BPSN1direct},
\end{eqnarray}
with $\epsilon_{L,R} = \frac12(1\pm \gamma_5)\epsilon$, and with
composite gauge field given by:
\begin{equation}
  A_\mu ^B=\ft{1}{2 
  }\rmi\left[ (\partial _i {\cal K})\partial _\mu \phi^i-(\partial
  ^i {\cal K} )\partial _\mu \phi_i\right]
  +
  W_\mu ^\alpha {\cal P}_\alpha .  \label{U1connection}
\end{equation}

The comparison with the $N=2$ formulae goes by the substitutions
\begin{align}
\epsilon _L  & =  \epsilon^1 ,\qquad &\psi _{\mu L} &= \psi_\mu^1,\nonumber\\
\lambda _L^\alpha   & =  -\lambda _2^\beta  {\cal D}_\beta  X^I, \qquad &
D^\alpha &=-2g\Im{\cal N}^{-1|IJ}{\cal P}^3_J,\nonumber\\
\Re f&=-\Im{\cal N},\qquad &{\cal P}&=2g{\cal P}^3.
 \label{compareN21}
\end{align}
Note that in the second line $\beta$ on the right-hand side is the index
related to coordinates of the special K{\"a}hler manifold, while the $\alpha
$ on the left-hand side labels the vectors, and corresponds to the index
$I$ on the right-hand side.

In the models that we consider we have $\Im{\cal N}_{IJ}=-\delta _{IJ}$.
This implies the relation (\ref{DinP3}). As in the point $z=0$ of the
special manifold that we consider,  (\ref{cKmodel}) leads to
\begin{equation}
  {\cal D}_zX^1=\rme^{{\cal K}/2} =\frac{1}{\sqrt{2}},
 \label{calDX1}
\end{equation}
the $N=1$ gaugino is
\begin{equation}
  \lambda _L = -\frac{1}{\sqrt{2}}\lambda _2.
 \label{lambdaLinlambda2}
\end{equation}
Therefore, the gravitino and gaugino transformation in (\ref{N=1BPS}) are
in agreement with (\ref{BPSN1direct}).

Identifying $\Phi $ with $\phi _i$, the kinetic terms are derivable from
a K{\"a}hler potential
\begin{equation}
  {\cal K}_{N=1}= -\alpha ^2\log(2\Im\Phi ).
 \label{KN1}
\end{equation}
This identifies also (\ref{ABfromN2}) with (\ref{U1connection}).

The case of one vector and one chiral multiplet with complex field $\Phi$
parametrizing the upper half plane $\frac{\SU(1,1)}{\U(1)}$ can now be
analysed. If the compact isometry (isotropy w.r.t. the base point $\Phi =
\rmi$) $\delta \Phi =2g(\Phi^2+1)$ is gauged and a constant FI term is
added, the corresponding $D$-term is obtained in (\ref{P3general}). The
vacuum manifold is a circle with centre at $(0,\rmi\xi/\alpha ^2)$ and
radius $(\sqrt{\xi^2/\alpha ^4-1})$.

If we now perform the gauging described above, and take the gauge
coupling function to be the identity, we see that the BPS equations
reduce to the equations (\ref{N=1BPS}) for $\epsilon^1$ that we get from
the truncation, as should be the case.

\subsection{Energy of $D$-term strings with general K{\"a}hler target spaces}
\label{sss:BogKah}

We determine the energy per unit length for cosmic string configurations
in the presence of an arbitrary number of chiral multiplets with a
generic K{\"a}hler-potential ${\cal K}$ and Abelian gauging (i.e. for a
generic choice of Killing vector $ \delta_\alpha \phi_i =-k_{\alpha i}
$.).

Once we take the ansatz for the metric as in (\ref{ansatzmetric}) and we
fix the field strength to have only non zero $F^\alpha_{r\theta}$
component, corresponding to a magnetic field in the $z$ direction,
we obtain the projector
\begin{equation}
  \gamma^{12} \epsilon_{L} = \mp
\rmi\epsilon_{L},\label{proje}
\end{equation}
and the following BPS conditions:
\begin{eqnarray}
   &   & \left(\nabla_r \pm \frac \rmi C
\nabla_\theta\right)\phi_i=0, \nonumber\\
   &   &F^\alpha_{12} \mp D^\alpha=0, \nonumber\\
    & &C''\pm
F^B_{r\theta}=0,
 \label{BPSeqns}
\end{eqnarray}
where $F^B_{\mu\nu}\equiv
\partial_\mu A^B_\nu -\partial_\nu A^B_\mu$.

The total energy per unit length is given by:
\begin{eqnarray}
{\cal \mu}_{\rm string}&=& \int \,\sqrt{\det g}\,\, \rmd r \rmd\theta
\left[ {g_i}^j(\nabla_\mu \phi_j)(\nabla^\mu \phi^j) + \frac{1}{4}(\Re
f_{\alpha \beta})
 F^\alpha_{\mu \nu }\, F^{\beta \mu \nu} +\frac{1}{2} D^2 -
 \frac{1}{2}
 R\right] \nonumber\\
&&+\left. \left( \int \rmd\theta \sqrt{\det h}\,\,K\right|_{r=\infty
}-\left.\int \rmd\theta \sqrt{\det h}\,\,K\right|_{r=0}\right) ,
 \label{estring}
\end{eqnarray}
where the sums over $\mu ,\nu $ run only over $r,\theta $. The quantity
$K$ is the Gaussian curvature at the boundaries (on which the metric is
$h$), which are at $r=\infty $ and $r=0$. For the
metric~(\ref{ansatzmetric}) we have:
\begin{equation}
\sqrt{\det g}=C(r),\qquad \sqrt{\det g}\,R=-2C'',\qquad \sqrt{\det h}\,
K=-C'.
\end{equation}

We now argue that the energy (\ref{estring}) can always be obtained in
terms of the BPS equations (\ref{BPSeqns}). Indeed, consider the
combination
\begin{eqnarray}
{\cal \mu}_{\rm string}=&&\int \, \rmd r\rmd\theta\, C(r)\left[  \,
 {g_i}^j(\nabla_r  \, \pm \, {\rm i}C^{-1} \, \nabla_\theta ) \phi^i
 (\nabla_r  \, \mp \, {\rm i}C^{-1} \, \nabla_\theta ) \phi^j\right. \nonumber\\
 &&\left. \qquad \, + \,
\ft12 \left(F^\alpha_{12} \, \mp   D^\alpha \right)(\Re f_{\alpha
\beta}) 
(F^\beta_{12} \, \mp D^\beta 
) \right]  \, \cr
 &&+  \int \rmd r \rmd\theta \,  \left[C''\pm F^B_{r\theta}
\right]- \left.\int \rmd\theta\, C'\right|_{r=\infty }+ \left.\int
\rmd\theta\, C'\right|_{r=0}.
\end{eqnarray}
Using (\ref{defPN1}), we can derive from (\ref{U1connection}):
\begin{equation}
 F^B_{\mu\nu} = 2\rmi {g_i}^j
\partial_{[\mu} \phi_j \partial_{\nu]} \phi^i +  F_{\mu\nu}^\alpha {\cal P}_\alpha
-2\rmi W^\alpha_{[\mu}\left[( \partial_{\nu]} \phi^i {k_\alpha}_j +
\partial_{\nu]} \phi_j {k_\alpha}^i) {g_i}^j
\right].
 \end{equation}
Now one can, after some calculation, reconstruct the kinetic terms for
the gauge field and the scalars, and the potential for the scalars in
(\ref{estring}).

The conclusion is that any $\U(1)$ gauging of a compact isometry on any
K{\"a}hler target space can give rise to a cylindrically symmetric BPS string
with mass per unit length given by the Gibbons-Hawking surface term (if
the BPS equations can be solved\footnote{An explicit numerical solution
of BPS equations in a similar situation has been given in
\cite{Davis:2005jf,Blanco-Pillado:2005xx}.}). We use this result when we
give the energy density for our string solution of section
\ref{ss:resultN1}. One can also check explicitly that the $N=1$ equations
of motion hold in this case. The analysis also applies to the semilocal
string solutions discussed in \cite{Urrestilla:2004,Dasgupta:2004dw} (see
also \cite{Chen:2005ae,Edelstein:1996ue}), and the axionic D-term strings
discussed in \cite{Blanco-Pillado:2005xx}, where a non trivial K{\"a}hler
potential was also considered.

However suggestive, this Bogomol'nyi-type argument cannot be used
directly to conclude that the solutions are stable, as non-axisymmetric
or $z$-dependent perturbations might destabilize the strings. But we
expect the Bogomol'nyi bound can be  generalized to non-axisymmetric and
multi-vortex configurations along the lines of \cite{Gibbons:1992gt}.

Also, when the BPS strings are non-topological, the presence of
cylindrically symmetric zero modes that make the magnetic field spread
can prevent the strings from forming in a cosmological context (see
\cite{Leese:1992hi,Benson:1993at,Achucarro:1998ux,Dasgupta:2004dw} for a
discussion of this point in the context of BPS semilocal strings).

\section{Discussion}
\label{ss:conclusions}

In this work, we have studied the embedding of four-dimensional $N=1$
$D$-term string solutions in $N=2$ supergravity models. We have shown how
an $N=2$ action with $\U(1)$ gauging can be reduced by a consistent
truncation to an $N=1$ action with Fayet-Iliopoulos term and vanishing
superpotential. Especially important in the construction are the choices
of gauging and special geometry.

Alternatively, one can use this information to devise an ansatz which
allows the full $N=2$ BPS equations to be solved explicitly. In this way
the half-BPS $N=1$ $D$-term strings are promoted to half-BPS $N=2$
strings.

The reduced $N=1$ action has a charged scalar parametrizing a non trivial
K{\"a}hler-Hodge target space. We have shown how to solve the resulting $N=1$
BPS-equations for this system, along the lines of \cite{Dvali:2003zh},
and demonstrated that the half-BPS solution is also half-BPS in the $N=2$
context. A full stability analysis of these solutions has not been
presented and is a very important issue. Another interesting related
question is the possible interactions of the strings with black holes.

The way to obtain arbitrary FI constants in $N=1$ by consistent
truncations can be useful in a wider context than that of string
solutions, while the results for $N=1$ theories with arbitrary
K{\"a}hler-Hodge geometries can be of interest in the current research
concerning cosmic string solutions in string theory.

We have presented our results for both normal quaternionic manifolds of
dimension one. The generalization to normal higher dimensional
quaternionic manifolds is straightforward since any normal quaternionic
manifold admits one of the quaternionic manifolds of dimension one as a
completely geodesic submanifold with a compatible quaternionic structure.

The issue of K{\"a}hler anomalies in gauged $N=1$ supergravity with a
non-trivial K{\"a}hler manifold was recently brought up in
\cite{Freedman:2005up}. The possible anomalies discussed in that paper are due to
the $\U(1)$ R-symmetry group. The setting in this paper of $N=2$ theories
avoids such problems. The R-symmetry group of $N=2$, $D=4$ supergravity
is $\U(1)\times \SU(2)$. The gauging that we consider is an Abelian
gauging that does not act on the scalars of the vector multiplets, but
only on the hyperscalars. As the hyperscalars are inert under the $\U(1)$
factor of the R-symmetry group, we are only concerned with the $\SU(2)$
factor. The hyperini transform under $\Sp(m,\mathbb{R})$, which is a real
representation and therefore free of anomalies. The gaugini and the
gravitini are charged under the $\SU(2)$ of the R-symmetry of $N=2$ and
transform as a doublet. But $\SU(2)$ is not an anomalous group for local
gauging so the gaugini and gravitini will not bring about any anomalies.
We can thus conclude that our gauging is going to be free of anomalies as
the fermions of the theory transform in an anomaly free representation
and the $\U(1)$ of the R symmetry (coming from special geometry) is not
gauged. Note that this  explanation is quite general and can apply to any
Abelian gauging in $N=2$ supergravity. From this point of view it is
safer to consider the string solution as living in $N=2$ although the
solution involves only fields related to an $N=1$ subsector.

BPS solitons and defects can usually be coupled to gravity without losing
their BPS character. This is true for $N=1$ $D$-term strings, as
\cite{Dvali:2003zh} showed. But in $N=2$ global supersymmetry there are
also half-BPS cosmic strings \cite{Achucarro:2004ry}, and the question
arises as to what is the fate of these solutions when coupled to
(super)gravity.  One would naively expect to be able to find the
corresponding half-BPS solutions in $N=2$ supergravity and moreover they
should not be too different from the $N=1$ ones, because in the absence
of gravity they are equivalent.  Until now, the stumbling block in giving
these solutions an $N=2$ interpretation was the difficulty in
constructing constant Fayet-Iliopoulos terms in $N=2$ supergravity and it
is reassuring that this difficulty can be circumvented.

\medskip
\section*{Acknowledgments.}

\noindent We are grateful to Gary Gibbons, Luca Martucci, Jorge Russo,
Kepa Sousa, Mario Trigiante and Stefan Vandoren for very useful
discussions. We also thank Gary Gibbons for making available to us
unpublished notes on gravitating topological defects.

This work is supported in part by the European Community's Human
Potential Programme under contract MRTN-CT-2004-005104 `Constituents,
fundamental forces and symmetries of the universe'. The work of A.C., J.
V.d.B. and A.V.P. is supported in part by the FWO - Vlaanderen, project
G.0235.05 and by the Federal Office for Scientific, Technical and
Cultural Affairs through the "Interuniversity Attraction Poles Programme
-- Belgian Science Policy" P5/27. The work of A.A. is supported in part
by the Netherlands Organization for Scientific Research (N.W.O.) under
the VICI Programme, and by the Spanish Ministry of Education through
projects FPA2002-02037 and FPA2005-04823.

\newpage
\appendix
\section{Notation}\label{app:notation}
Our metric is mostly $+$, and we use the $(+++)$ conventions in the
Misner-Thorne-Wheeler classification \cite{MisnerThorneWheeler} scheme,
such that compact spaces have positive scalar curvature, and covariant
derivatives on fermions have the form
\begin{equation}
  \partial _\mu +\ft14 \omega_\mu{}^{ab}\gamma _{ab}.
 \label{signomegaspinors}
\end{equation}
Antisymmetrization is done with weight~1, see (\ref{defF+-}). We use
indices $\mu $ and $a$ for local and tangent spacetime. The $N=2$
extension index is $i=1,2$. This is related to $\SU(2)$ vectors, labelled
by $x=1,2,3$, using the Pauli matrices:
\begin{equation}
A_i{}^j\equiv \rmi  A^x  \left( \sigma^x\right) _i{}^j, \qquad
\mbox{or}\qquad A^x=-\ft12\rmi \tr \sigma ^x A.
\end{equation}
Lowering and raising SU(2) indices is done using the $\varepsilon$
symbol, in northwest-southeast (NW-SE) conventions,
\begin{equation}
A^i=\varepsilon^{ij}A_j, \qquad A_i=A^j\varepsilon_{ji}.
\end{equation}
$\gamma _5$ and the Levi-Civita symbol are normalized as
\begin{equation}
\varepsilon_{0123}=1,\qquad \varepsilon^{0123}=-1,\qquad \gamma_5\equiv
\rmi \gamma_0\gamma_1\gamma_2\gamma_3. \label{valueLeCi}
\end{equation}

Quaternions are written as $2\times 2$ complex matrices using
\begin{equation}
  q=q^0 \sigma^0 +\rmi  q^x  \sigma^x.
 \label{quaternion22}
\end{equation}
Hermitian conjugation and the anti-Hermitian part are denoted as
\begin{equation}
\bar q=q^0 \sigma^0 -\rmi  q^x  \sigma^x,\qquad
  \vec q = \rmi q^x \sigma^x.
 \label{imagquat}
\end{equation}

Translations between forms and components are done with factors and signs
as in
\begin{equation}
  J=\ft12 J_{XY}\rmd q^X\rmd q^Y,\qquad \rmd (A_X\rmd q^X)= \partial
  _YA_X \rmd q^Y\rmd q^X.
 \label{conventionsforms}
\end{equation}

All the properties and conventions on hypermultiplets that we follow can
be found in \cite{Bergshoeff:2002qk}, especially in appendix~B, except
from the change of notation that we now indicate a 3-vector with indices
$x$ rather than $\alpha $ in that paper. The spinors of hypermultiplets
are labelled by $A=1,\ldots ,2n_H$. These are $\Sp(n_H)$ indices, which
we will sometimes split in further $\SU(2)$ indices $i =1,2$ and vector
indices $t,s=1,\ldots ,n_H$. The former index will be put in opposite
up/down position, such that e.g. the 1-form vielbein $f^{iA}$ becomes
with $A=(tj)$ for every value of $t$ a $2\times 2$ matrix $f^t{}_j{}^i$.
The symplectic metric is then split as
\begin{equation}
  \mathbb{C}_{AB}=\varepsilon ^{ij}\otimes \unity _{ts} \qquad \mbox{for
  }A=(ti),\,B=(sj).
 \label{splitSymplMetric}
\end{equation}
The reality condition (with complex conjugation denoted as $*$),
\begin{equation}
  (f^{iA})^*= f^{jB}\varepsilon_{ji}\mathbb{C}_{BA},
  \label{realVielbein}
\end{equation}
translates then to the property that $f^t$ as a $2\times 2$ matrix
(`quaternion') satisfies
\begin{equation}
  (f^t)^*=\sigma _2f^t\sigma _2.
 \label{realfmatrix}
\end{equation}
This is satisfied for quaternions of the form (\ref{quaternion22}) with
$q^0$ and $q^x$ real. We have then
\begin{eqnarray}
 && g_{XY}f^Y_{iA}=(\bar f_X^t)_i{}^j\quad \mbox{with }A=(tj), \qquad g_{XY}=\tr(f_X^t \bar f_Y^t),\nonumber\\
&&  (J^x)_{XY}=(J^x)_X{}^Zg_{ZY}=-\rmi \tr(f^t_X\sigma ^x\bar f^t_Y),
 \label{quatgJ}
\end{eqnarray}
or the 2-form hypercomplex structure as a quaternion is
\begin{equation}
  \vec{J}=-\bar f^t\wedge f^t.
 \label{hypercomplexmatrix}
\end{equation}

\section{Parametrization of coset spaces}\label{app:paramCosets}
We encounter in this paper two 1-dimensional quaternionic-K{\"a}hler  coset
spaces. Both can be expressed as submanifolds of the 2-dimensional
quaternionic-K{\"a}hler manifold $\frac{\Sp(2,1)}{\Sp(2)\times \Sp(1)}$. We
give here the parame\-trizations that we use. Though the 1-quaternion
coset spaces can be used without reference to the 2-quaternion one, we
will start by the parametrization of the latter, and determine
parametrizations of the others as truncations thereof.
\subsection{Parametrization of the 2-dimensional projective quaternionic space}
We shall consider the quaternionic-K{\"a}hler manifold of quaternionic
dimension 2:
\begin{equation}
  \frac{\Sp(2,1)}{\Sp(2)\times \Sp(1)}\simeq
\frac{\USp(4,2)}{\USp(4)\times \USp(2)}.
 \label{Coset2}
\end{equation}
The algebra of the isometry group, $\fsp(2,1)$ can be defined as the set
of matrices over the quaternions $\mathbb{H}$ that preserve a metric of
signature $(+,+,-)$. We take this metric in the form
\begin{equation}
  \mu =\begin{pmatrix} & & 1 \\
& 1 & \\
 1 & & \end{pmatrix},
 \label{mumetricSp}
\end{equation}
where each entry is a quaternion, or $2\times 2$ complex matrix. The
elements  $M$ of $\fsp(2,1)$ are those $3\times 3$ matrices with entries
in $\mathbb{H}$ that satisfy
\begin{equation}
  \mu M^\dagger \mu = -M.
 \label{defnsp21}
\end{equation}

The general form of an element of $\fsp(2,1)$ is then
\begin{equation} M=
\begin{pmatrix}
a & \ft12(\bar e+\bar f)  & -\ft12(\vec b+\vec{c}) \\[1mm]
\ft12(e-f) & \vec p & -\ft12( e+f)  \\[1mm]
\ft12(\vec{b}-\vec c)  & \ft12(\bar f-\bar e) & -\bar a
\end{pmatrix},
\end{equation}
where $a=a_0+\vec a$, $e=e_0+\vec e$  and $f=f_0+\vec f$ are generic
quaternions  and $\vec c$, $\vec b$ and $\vec p$ are pure anti-Hermitian
quaternions (with vanishing Hermitian part).\footnote{The identification
$\fsp(2,1)\simeq \fusp(4,2)$ is obtained once we take the matrices  $
-\ii \vec \sigma $ for the imaginary quaternions.}

The Lie  algebra of $\fsp(2,1)$ can be split into  a compact (anti-Hermitian) and non-compact  (Hermitian) part :
\begin{equation} M_{H}=
 \begin{pmatrix}
\vec a & \ft12\bar f & -\ft12\vec c \\[1mm]
-\ft12 f & \vec p & -\ft12 f  \\[1mm]
-\ft12\vec c &  \ft12\bar f & \vec a
\end{pmatrix}
, \quad M_{G/H}=
 \begin{pmatrix}
 a_0 \unity  & \ft12\bar e & -\ft12\vec b \\
\ft12 e & 0  & -\ft12 e  \\ \ft12\vec b &  -\ft12\bar e & - a_0\unity
\end{pmatrix}.
\end{equation}
The $H$ part of the generator can be decomposed into its
subalgebras\footnote{It is related to the previous expression of $M_{H}$
by  taking $\vec u=\ft12\vec a+\ft14\vec c$  and $\vec v=\ft12\vec
a-\ft14\vec c$.} :
\begin{equation}  M_{\fsu( 2)}=
\begin{pmatrix}
\vec u  & 0 & -\vec u \\
0 & 0  & 0  \\
-\vec u &  0  &  \vec u
\end{pmatrix}
,\quad M_{\fsp(2)}=
\begin{pmatrix}
\vec v  & \ft12\bar f & \vec v \\[1mm]
-\ft12 f & \vec p  & - \ft12 f  \\[1mm]
\vec v  &  \ft12\bar f  & \vec v
\end{pmatrix}.
\end{equation}
$ M_{\fsp(1)}$ commutes with $M_{\fsp(2)}$ and the latter contains two
commuting $\fsu(2)$ parameterized by $\vec p$ and $\vec v$:
\begin{equation}
M_{\fsu(2)\oplus \fsu(2)\subset \fsp(2)}=
\begin{pmatrix}
\vec v  & 0 & \vec v \\
0       & \vec p       & 0         \\
\vec v  &  0  & \vec v
\end{pmatrix}.
\end{equation}
We see that the compact subalgebra of $\fsp(2,1)$ contains  three
commuting $\fsu(2)$. $M_{\fsu(2)}\subset\fsp(1)$ corresponds to the
R-symmetry whereas the $\fsu(2)_{\vec p}\subset \fsp(2) $  contains the
compact $\rU(1)$ for the string.

The solvable gauge of the coset manifold is obtained  by adding to
$M_{G/H}$ an element of $M_H$ (with $\vec c=\vec b$, $f=e$ and $\vec
a=\vec p=0$) so that the result is an upper triangular matrix:
\begin{equation}
M_{\rm Solvable}=
 \begin{pmatrix}
 a_0 \unity & \bar e & -\vec b \\
0 & 0  & - e  \\
0 &  0 &
- a_0 \unity
\end{pmatrix}.
\end{equation}

\subsection{Solvable coordinates and metric of $\frac{\Sp(2,1)}{\Sp(2)\Sp(1)}$}
\label{ss:coordSp21}

 We parametrize the coset elements by
\begin{equation}
L=\rme^N \cdot \rme^H,
\end{equation}
where
\begin{equation} N=N_e+N_b=
\underbrace{
\begin{pmatrix}
0 & \bar e & 0\\
0 & 0 & - e \\
0 & 0 & 0
\end{pmatrix}
}_{N_e}
+
\underbrace{
\begin{pmatrix}
0 & 0 & -\vec b \\
0 & 0 & 0 \\
0 & 0 & 0
\end{pmatrix}
}_{N_b}
\, ,\quad
H=
\frac{1}{2}
\begin{pmatrix}
h \unity  & 0 & 0 \\
0 & 0 & 0 \\
0 & 0 & -h \unity
\end{pmatrix}.
\end{equation}
The coordinates $q^X$ are thus the real $h$, the 3 real coordinates of
$\vec{b}$ and the 4 real parts of the quaternion $e$. This leads to
\begin{equation} L=
 \begin{pmatrix}
\rme^{\frac{1}{2}h} \unity & \bar e & -\rme^{-\frac{1}{2}h}(\vec b +\frac{\bar e e }{2}) \\
0 & \unity & - \rme^{-\frac{1}{2}h} e \\
0 & 0 & \rme^{-\frac{1}{2}h}\unity
\end{pmatrix}.
\end{equation}
This leads to the algebra element
\begin{equation} L^{-1} \rmd L=
\begin{pmatrix}
\frac{B_0}{2} & \frac{\bar E}{\sqrt{2}} & -\vec B \\
0 & 0 & -\frac{E}{\sqrt{2}} \\
0 & 0 & -\frac{B_0}{2}
\end{pmatrix},
\end{equation}
where
\begin{equation}
B=B_0\unity+\vec B= \rmd h  \unity +\rme^{-h}\left[\rmd\vec b
-\ft{1}{2}(\bar e \rmd e-\rmd \bar e e )\right]\, ,\quad  E= \sqrt{2}\,
\rme^{-\frac{1}{2}h} \rmd e, \label{defBE}
\end{equation}
or in real components
\begin{equation}
B_0=\rmd h,\qquad B^x=\rme^{-h}\left({ \rmd b^x+  e^x \rmd e^0-e^0 \rmd
e^x- \varepsilon^{xyz} e^y \rmd e^z   }\right).
\end{equation}
The algebra element can be split in the coset part and the part in $H$.
The first one is the Hermitian part:
\begin{equation}
(L^{-1} \rmd L)_{G/H}= \frac{1}{2}
\begin{pmatrix}
B_0 & \frac{ \bar E}{\sqrt{2}} & -\vec B \\
\frac{ E}{\sqrt{2}} & 0 & -\frac{ E}{\sqrt{2}} \\
\vec B & -\frac{\bar E}{\sqrt{2}} & - B_0.
\end{pmatrix}.
\end{equation}
The part in $H$ is the anti-Hermitian part, which can be split in the
$\fsp(1)$ and $\fsp(2)$ part:
\begin{eqnarray}
(L^{-1} \rmd L)_{H}&=& \frac{1}{2}
\begin{pmatrix}
0 & \frac{\bar  E}{\sqrt{2}} & -\vec B \\
-\frac{E}{\sqrt{2}} & 0 & -\frac{E}{\sqrt{2}} \\
-\vec B & \frac{\bar E}{\sqrt{2}} & 0
\end{pmatrix}
=(L^{-1} \rmd L)_{\fsp(1)}+(L^{-1} \rmd L)_{\fsp(2)},
\nonumber\\
&&(L^{-1} \rmd L)_{\fsp(1)}= \frac{1}{4}
\begin{pmatrix}
 \vec B & 0 & -\vec B \\
0 & 0 & 0 \\
-\vec B & 0 & \vec B
\end{pmatrix}, \nonumber\\
&&(L^{-1} \rmd L)_{\fsp(2)} =
\begin{pmatrix}
-\frac{1}{4}\vec B & \frac{\bar  E}{\sqrt{2}} &-\frac{1}{4} \vec B \\
-\frac{ E}{\sqrt{2}} & 0 & -\frac{E}{\sqrt{2}} \\
-\frac{1}{4}\vec B & \frac{\bar E}{\sqrt{2}} & -\frac{1}{4}\vec B
\end{pmatrix}.
\end{eqnarray}
The metric is defined as
\begin{equation}
\rmd s^2=g_{XY}\rmd q^X\rmd q^Y  =  \Tr \left[{ (L^{-1} \rmd
L)_{G/H}\cdot (L^{-1} \rmd L)_{G/H} }\right] = \ft{1}{2}\tr ( B\bar B+
E\bar E ), \label{ds2def}
\end{equation}
where $\Tr$ stands for a trace over the $6\times 6$ matrix and $\tr$ for
a trace over the $2\times 2$ matrix. We will comment on the normalization
of this metric below. Its value is
\begin{equation}
\rmd s^2= (\rmd h)^2+ (B^1)^2+(B^2)^2+(B^3)^2 +2\rme^{-h}\left[{ (\rmd
e^0)^2+(\rmd e^1)^2+(\rmd e^2)^2+(\rmd e^3)^2 }\right].
\label{metricSp21}
\end{equation}
The vielbeins, as 1-forms and quaternions as explained above, can be
taken to be
\begin{equation}
  f^1= \frac{1}{\sqrt{2}} B,\qquad f^2 =\frac{1}{\sqrt{2}} E.
 \label{f1f2}
\end{equation}
These lead to (\ref{ds2def}) and to the hypercomplex form ($\wedge $
symbols understood)
\begin{equation}
  \vec{J}=-\ft12 \left(\bar B\,  B+\bar E\,  E\right),\qquad \mbox{or}\qquad
  J^x=-B_0B^x -E_0 E^x -\ft12\varepsilon ^{xyz}\left( B^yB^z
  +E^y E^z\right).
 \label{vecJmatrix}
\end{equation}
Using the differentials
\begin{eqnarray}
  &&\rmd B= -B_0  B-\ft12\bar E\, E, \qquad \rmd E=-\ft12B_0
  E,\nonumber\\
   &&\mbox{or}\qquad \rmd B^x= -B_0 B^x-E_0E^x-\ft12\varepsilon
  ^{xyz} E^yE^z,
 \label{dBdE}
\end{eqnarray}
we obtain
\begin{equation}
  \rmd J^x +2\varepsilon ^{xyz}\omega ^yJ^z=0,
 \label{dJomega}
\end{equation}
for
\begin{equation}
  \omega ^x=-\ft12 B^x.
 \label{valueomega}
\end{equation}
We find then that (\ref{RSU2isJ}) is satisfied for $\nu =-1$. The value
that we get here for $\nu $ depends on the normalization of the metric.
Multiplying the metric by an arbitrary $-\nu^{-1}$, would lead to
(\ref{RSU2isJ}) with this arbitrary value of $\nu $. In the supergravity
context, $\nu = -\kappa ^2$, where $\kappa $ is the gravitational
coupling constant, which we have put equal to~1.

\subsection{The projective quaternionic manifold $\frac{\rSp(1,1)}{\rSp(1)\rSp(1)}$}

The quaternionic-K{\"a}hler manifold of quaternionic dimension 1
\begin{equation}
 \frac{\rSp(1,1)}{\rSp(1)\rSp(1)}
\end{equation}
can be seen as a submanifold of $\frac{\rSp(2,1)}{\rSp(2)\rSp(1)}$ by
taking $E=0$. The metric is then
\begin{equation}  \rmd s^2=
(\rmd h)^2+ \rme^{-2h}\left[{ (\rmd b^1)^2+(\rmd b^2)^2+(\rmd
b^3)^2}\right].
 \end{equation}
The vielbein and hypercomplex forms are simply obtained from the previous
section by putting $e=E=0$. E.g.
\begin{equation}
{\cal R}^x= \ft12{\rme^{-h}}\rmd h \wedge \rmd b^x+\ft14
\rme^{-2h}\varepsilon ^{xyz} \rmd b^y\wedge \rmd b^z , \quad x,y,z=1,2,3,
\end{equation}
and can be obtained as the curvature of the connection
\begin{equation}
\omega^x=-\ft12\rme^{-h} \rmd b^x.
\end{equation}

\subsection{The normal quaternionic manifold $\frac{\rSU(2,1)}{\rU(2)}$}
 \label{app:SU21}
The quaternionic-K{\"a}hler manifold of dimension one
\begin{equation}
\frac{\rSU(2,1)}{\rU(2)}
\end{equation}
can be defined as a submanifold of $\frac{\rSp(1,1)}{\rSp(1)\rSp(1)}$ in
many different ways. Here we will consider the choice
\begin{equation}
b^1= b^2=e^0 =e^3=0.
\end{equation}
The metric can be obtained from reducing (\ref{metricSp21}). However, in
order to respect the $\nu =-1$ normalization as explained at the end of
section \ref{ss:coordSp21}, we need here another overall factor. We have
\begin{equation}
\rmd s^2= \ft12\rmd h^2+\ft12\rme^{-2h}(\rmd b^3-e^1 \rmd e^2+e^2 \rmd
e^1)^2+ \rme^{-h}\left[(\rmd e^1)^2+(\rmd e^2)^2\right].
\end{equation}
We can again obtain all expressions from those of
$\frac{\rSp(1,1)}{\rSp(1)\rSp(1)}$, where $B$ has only the $B_0$ and
$B^3$ components, and $E$ has only $E^1$ and $E^2$. The vielbein is
\newcommand{\vs}{v}
\newcommand{\ut}{t}
\begin{equation}
  f=\frac{1}{2}(\bar B+\bar E)=\begin{pmatrix}
\vs & - \ut^* \\
\ut & \vs^*
\end{pmatrix},
 \label{vierbeinnormal}
\end{equation}
where
\begin{equation}
 \vs=\frac{ 1}{2}\left[{\rmd h-\ii \rme^{-h}(\rmd b^3-e^1 \rmd e^2+e^2 \rmd e^1)}\right],
  \quad \ut=\frac{1}{\sqrt{2}}\rme^{-h/2}(\rmd e^2-\ii \rmd e^1).
\end{equation}
The quaternionic form is
\begin{equation}
{\cal R}=\rmi{\cal R}^x (\sigma^x)= \frac12 \bar f\wedge f=\frac{1}{2}
  \begin{pmatrix}
 \vs^*\wedge \vs + \ut^*\wedge \ut & 2\, \ut^*\wedge \vs^* \\
  2\vs  \wedge \ut &  \vs\wedge \vs^*+  \ut\wedge \ut^*
  \end{pmatrix},
\end{equation}
which gives
\begin{eqnarray}
{\cal R}^1 &=&\frac{1}{2\sqrt{2}} \left[\rme^{-h/2} {\rmd e^1\wedge \rmd
h+ \rme^{-3h/2} \rmd e^2 \wedge  (\rmd b^3+e^2 \rmd e^1)\, }\right], \nonumber\\
{\cal R}^2 &=&\frac{1}{2\sqrt{2}} \left[\rme^{-h/2}{\rmd e^2\wedge \rmd h
- \rme^{-3h/2}\rmd e^1 \wedge  (\rmd b^3-e^1 \rmd e^2)\, }\right], \nonumber\\
{\cal R}^3 &=& -\frac14\rme^{-h} \rmd h \wedge(\rmd b^3+e^2 \rmd e^1-e^1
\rmd e^2)+ \frac{1}{2}\rme^{-h} \rmd e^1\wedge \rmd e^2,
\end{eqnarray}
and is the curvature of the connection
\begin{equation}
\omega^1=\frac1{\sqrt{2}} \rme^{-h/2}\rmd e^1,\quad  \omega^2=
\frac1{\sqrt{2}} \rme^{-h/2}\rmd e^2 ,\quad \omega^3=\frac14\rme^{-h}
(\rmd b^3+e^2 \rmd e^1-e^1 \rmd e^2).
\end{equation}

This parametrization of the manifold can also be found in
\cite{Ceresole:2001wi} with the following replacements:
\begin{equation}
  h=\log V,\qquad b^3=-\sigma ,\qquad  e^1=-\sqrt{2}\tau ,\qquad
  e^2=\sqrt{2}\theta.
 \label{conventionCeresole}
\end{equation}
The chosen vielbeins differ by a multiplication by $\rmi\sigma _2$ on the
side of the indices $A,B$. That does not change the complex structures.
Note, however, that our conventions differ by the two signs in
(\ref{conventionsforms}) such that the 2-forms differ by a sign.

\providecommand{\href}[2]{#2}\begingroup\raggedright\endgroup

\end{document}